\documentclass[12pt]{article}

\usepackage[margin=1in]{geometry}
\usepackage{setspace}
\usepackage[flushleft]{threeparttable}
\usepackage[dvipsnames]{xcolor}
\usepackage{fancyhdr}

\usepackage{amsmath}
\usepackage{amssymb}
\usepackage{amsfonts}
\usepackage{bm}
\usepackage{resizegather}
\DeclareMathOperator{\E}{\mathbb{E}}

\usepackage{graphicx}
\usepackage{subcaption}
\usepackage{caption}
\usepackage{adjustbox}
\usepackage{tabularx}
\usepackage{booktabs}
\usepackage{rotating}
\usepackage{lscape}
\usepackage{makecell}

\usepackage{natbib}
\usepackage[colorlinks,citecolor=blue,urlcolor=blue]{hyperref}

\usepackage{amsthm}
\newtheorem{definition}{Definition}

\newtheorem{assumption}{Assumption}
\newtheorem{lemma}{Lemma}
\newtheorem{corollary}{Corollary}
\newtheorem{theorem}{Theorem}
\newtheorem{proposition}{Proposition}

\newtheorem{remark}{Remark}

\usepackage{enumitem}

\usepackage[all]{nowidow}
\usepackage[splitrule,flushmargin,bottom]{footmisc}

\title{Measuring Hidden Consumer Heterogeneity with Revealed Preferences}

\author{Avner Seror\thanks{avner.seror@univ-amu.fr. Aix Marseille Univ, CNRS, AMSE, Marseille, France. I am grateful to Yann Bramoull\'{e}, Pierre Bertrand, Yujian Chen, Laurens Cherchye, Sam Cosaert, Pawel Dziewulski, Joshua Lanier, Tom Potoms, John Quah, John Rehbeck, Reha Tuncer, Thierry Verdier, and the audience at the SAET 2025 conference for their comments and suggestions. All errors are my own. I acknowledge funding from the French government under the ANR
JCJC “BIAS" project (reference: ANR-25-CE26-7164-01) and the “France 2030” investment plan managed by the French National Research Agency (reference: ANR-17-EURE-0020) and from the Excellence Initiative of Aix-Marseille University – A*MIDEX. Mistakes are my own.}}
\date{June 2026}

\begin{document}

\maketitle

\begin{abstract}
Consumer heterogeneity in revealed-preference data is larger than bilateral rationality tests can reveal. We construct a continuous nonparametric metric of this hidden heterogeneity by repeatedly subsampling choices, partitioning agents into groups whose pooled data are jointly rationalisable under a chosen consistency criterion and recording how often each pair is co-classified. The resulting kernel is positive semi-definite, embeds the population in a Hilbert space, and induces a metric with the triangle inequality. Under a necessary-and-sufficient contrast-rank condition, its spectral structure recovers latent preference types. Inference on demographic correlates proceeds via a Monte-Carlo-conditional test and a finite-sample-valid permutation test. Applied to US grocery scanner data, the construction reveals a joint-rationality gap of $0.62$ between near-saturated pairwise compatibility and population-level co-typing; binary lottery data yield a comparable gap of $0.38$. Standard demographics organise only a modest part of the scanner kernel structure.
\end{abstract}


\clearpage

\section{Introduction}\label{sec:intro}

Consumer heterogeneity in revealed-preference data is larger than bilateral rationality tests can reveal. Pairwise GARP tests classify two consumers as compatible whenever the data lack the budget variation to distinguish them \citep{bronars}, and in many real datasets they saturate near full compatibility regardless of how different consumers actually are. Applied questions about how much consumers actually differ have therefore largely migrated to parametric demand systems, where distance is read off the parameter space. What has been missing on the revealed-preference side is a way to quantify the heterogeneity that pairwise tests leave hidden. The methodological gap is the absence of a population-endogenous metric of revealed-preference compatibility - a distance that can be read off the data without imposing a functional form.

We construct such a metric by repeated subsampling and partition. From a panel of consumer choices, we draw reduced subsamples of each agent's decisions, partition the population into groups whose pooled choices are jointly rationalisable, and record for each pair of agents the frequency with which they are co-classified across draws. The resulting similarity matrix is positive semi-definite, so the theory of positive definite kernels \citep{aronszajn1950} embeds the population in a Hilbert space and induces a distance satisfying the triangle inequality - which we prove both via the Hilbert norm and via a direct combinatorial argument exploiting the transitivity of block membership. The economic content of the distance comes from how the partition exploits this transitivity. A pairwise check between two agents detects only direct conflicts between them; the partition amplifies this by enforcing GARP at the block level, so that two agents are co-typed only if their choices are jointly rationalisable with every other agent in the block. By averaging co-typing events over many draws, the construction turns binary block membership into a continuous, population-endogenous measure of compatibility.

The mathematical structure of the construction - positive semi-definiteness, the Hilbert embedding, the triangle inequality of the induced distance - is unconditional. The kernel also admits a spectral decomposition into orthogonal axes of heterogeneity, ordered by their share of population variation. Under a necessary and sufficient contrast-rank condition on the type-level partition kernel, the leading axes recover the latent type structure: spectral clustering on the realised matrix achieves vanishing misclassification rate as the population and Monte Carlo draws grow. The contrast-rank condition is checkable when types are observed, and has a simple two-type interpretation: it reduces to ``average within-type co-typing exceeds cross-type co-typing.'' We frame the recovery result as an oracle-kernel recovery theorem rather than a primitive identification proof, conditional on the realised kernel approximating a stable type-level limit. The result extends to settings where individual consumers satisfy GARP only approximately - accommodating the small choice mistakes, measurement error, and short-run optimisation noise that are unavoidable in real choice data. We complement the recovery theory with two tests for whether observable demographics organise the kernel structure: an asymptotic Monte-Carlo-conditional test and a permutation test for which we establish finite-sample exactness and consistency whenever the demographic-type association is detectable through the kernel mean-difference contrast.

We apply the framework first in simulation. A controlled Cobb--Douglas environment with known preference types isolates the partition mechanism. As each agent contributes more observations per draw, same-type pairs are discriminated more sharply from cross-type pairs. The contrast-rank condition underwriting the recovery theorem is verified numerically, and the kernel cleanly separates structured rational behaviour from a Bronars-style random benchmark. Under continuous parameter heterogeneity, the number of observations each agent contributes per draw acts as a smoothing bandwidth: with few, the kernel measures similarity smoothly across nearby preferences; with many, it sharpens into a near-indicator of preference identity. The simulation establishes that the mechanism the theory predicts is the mechanism the kernel exhibits.

The first empirical application is to the \emph{Stanford Basket} scanner panel, used by \citet{bell1998}, \citet{shum2004}, \citet{hendel2006,hendel2006_b}, and \citet{echenique2011}, a $480$-household US grocery scanner panel of monthly purchases across $379$ food categories over $26$ months. Pairwise compatibility is near-saturated --- the mean pairwise GARP-consistency probability across household pairs is $0.99$ and $7\%$ of pairs have zero observation-level violations --- while only $37\%$ are jointly rationalisable when embedded in the population, leaving a joint-rationality gap of $0.62$. The kernel's spectral decomposition has a leading axis of heterogeneity that carries roughly three times the weight of the next, and family size is the most stable demographic correlate of where households sit on it. Headline demographic contrasts retain sign and rank across a $10\%$ Afriat-tolerance relaxation, so the conclusions about which demographics organise the kernel are not driven by strict GARP testing. 

The second is the Choice Prediction Competition 2018 dataset of $686$ subjects \citep{cpc18_dataset,erev2017}, with first-order stochastic dominance and the corresponding pooled-acyclicity condition replacing GARP. The same construction yields a joint-rationality gap of $0.38$ at one choice per subject, sharpening to $0.94$ at ten. As more choices per subject become available, the kernel's spectral structure resolves from a single dominant axis (initially four times stronger than the next) into multiple comparable axes of similar weight, consistent with the resolution of latent type structure that Theorem~\ref{thm:finite_type_witness} predicts. The lottery application is built to trace this progression across a wider range of within-subject sample sizes than is practical with the scanner panel.

Our work contributes to a literature on nonparametric approaches to preference heterogeneity \citep{gross95,crawford2012,heufer2014,castillo2018,cosaert2019,cherchye2024,Surana2022,Miao2025}. We build on \citet{crawford2012} but replace a single deterministic assignment with repeated subsampling, producing a continuous distance rather than a binary classification. The most thematically related contributions are \citet{cherchye2024}, who measure the partition-cardinality explained by a demographic and assess robustness through a subsampling stability check, and \citet{Surana2022}, who constructs bounds on the Kemeny distance between revealed-preference rankings to group individuals into types. Our framework differs in three ways. First, the object is a continuous PSD pairwise kernel with a Hilbert embedding and an induced population-level metric - not a scalar partition-cardinality summary or a pairwise-ranking distance - which is what enables the spectral decomposition and the joint-rationality-gap construction. Second, our resampling is at the within-agent observation level and constructs the pairwise co-typing probabilities; it does not assess stability of a population-level scalar. Third, we establish finite-sample exactness and consistency of the permutation test, complementing the permutation framework of \citet{chercye2023_approx_test} for testing approximate utility maximization against random consumer behaviour.


\section{Set-up and Definitions}\label{sec:setup}

We consider the standard consumer problem with $M\geq 2$ goods. A decision maker chooses a bundle $q\in \mathbb{R}^M_+$ subject to a linear budget constraint, where prices are given by a vector $p\in \mathbb{R}^M_{++}$.  Let $\mathcal{I}$ denote a finite set of agents. For each agent $i\in\mathcal I$, let $\mathcal{N}_i=\{1,\dots, N_i\}$ be the index set of observations and
\[
D_i=\{(p_i^n,q_i^n)\}_{n\in\mathcal N_i}
\]
the corresponding individual dataset. We write
\[
\mathcal{A}^n_i =\{q\in \mathbb{R}^M_+ : p_i^n\cdot q \leq p_i^n\cdot q_i^n\}
\]
for the choice set of agent $i$ at observation $n\in\mathcal N_i$. We maintain the assumption that each agent's preferences are stable across observations: the dataset $D_i$ records choices generated by a single, fixed preference relation (or equivalently, a single utility function) at different budget sets. If preferences change over time, the similarity measure would reflect the average compatibility across the mixture of preference states, potentially attenuating the detected heterogeneity. When no confusion arises, we drop the index $i$ from the notation.

\subsection*{Revealed Preference Conditions}

\begin{definition}\label{def:rationality}
For agent $i\in \mathcal{I}$, bundle $q^n$ is
\begin{enumerate}
    \item directly revealed preferred to bundle $q$, denoted $q^n R^0 q$, if $ p^n \cdot q^n \geq p^n \cdot q $;
    \item directly revealed strictly preferred to bundle $q$, denoted $q^n P^0 q$, if $ p^n \cdot q^n > p^n \cdot q $;
    \item revealed preferred to bundle $q$, denoted $q^n R q$, if there exists a sequence of observed bundles $q^j, \dots, q^m$ such that $q^n R^0 q^j R^0 \cdots R^0 q^m R^0 q$;
    \item revealed strictly preferred to bundle $q$, denoted $q^n P q$, if there exists a sequence of observed bundles $q^j, \dots, q^m$ such that $q^n R^0 q^j R^0 \cdots R^0 q^m R^0 q$, with at least one of the $R^0$ relations replaced by $P^0$.
\end{enumerate}
\end{definition}

\begin{definition}[GARP]\label{def:garp}
A dataset $D=(p^n,q^n)_{n\in\mathcal W}$ indexed by $\mathcal W$ satisfies the Generalized Axiom of Revealed Preference (GARP) if, for every finite sequence of observations $t_1,\dots,t_J$ in $\mathcal W$,
\[
q^{t_1} R^0 q^{t_2} R^0 \cdots R^0 q^{t_J} \;\;\Longrightarrow\;\; \text{not } (q^{t_J} P^0 q^{t_1}).
\]
\end{definition}

\begin{definition}[Rationalizability]\label{def:rationalizable}
A dataset $D=(p^n, q^n)_{n\in \mathcal{W}}$ is rationalizable if there exists a utility function $u: \mathbb{R}_+^M\rightarrow \mathbb{R}$ such that $u(q^n)\geq u(q)$ for all $q\in \mathcal{A}^n$.
\end{definition}

\begin{theorem}[Afriat, \citeyear{afriat1967}]\label{thm:afriat}
A dataset $D=(p^n, q^n)_{n\in \mathcal{W}}$ satisfies GARP if and only if there exists a strictly monotone, continuous, and concave utility function that rationalises $D$.
\end{theorem}

\begin{definition}[Partition]\label{def:partition}
Given a dataset $D=\{(p_i^n,q_i^n)\}_{i\in \mathcal{I},\, n\in \mathcal{N}_i}$, a partition of $\mathcal{I}$ is a collection $\Phi(D) = \{I_1, \dots, I_{L}\}$ with $\bigcup_{k=1}^{L} I_k = \mathcal{I}$ and $I_k \cap I_{k'} = \emptyset$ for all $k\neq k'$.
\end{definition}

\begin{definition}[GARP-consistent partition]\label{def:GARP_partition}
A partition $\Phi(D)= \{I_1, \dots, I_{L}\}$ is GARP-consistent if, for every block $I_k$, the pooled dataset $\{(p_i^n,q_i^n)\}_{i\in I_k,\; n\in \mathcal{N}_i}$ satisfies GARP.
\end{definition}

By Theorem~\ref{thm:afriat}, each block in a GARP-consistent partition is rationalizable by a strictly monotone, continuous, and concave utility function.

\section{Monte Carlo Subsampling}\label{sec:subsampling}

\subsection{Construction}

From a panel in which each agent $i$ is observed at $N_i$ budget sets, we construct a similarity measure by repeatedly drawing reduced datasets, partitioning agents into GARP-consistent types, and recording co-typing frequencies.

Each draw $t\in\{1,\dots,T\}$ proceeds as follows. For each agent $i\in\mathcal I$, sample $N$ observations (with replacement) from the $N_i$ available, producing a reduced dataset
\[
D^t=\bigl\{(p^{\,n_{i,s}^t}_i,\, q^{\,n_{i,s}^t}_i)\bigr\}_{i\in\mathcal I,\; s=1,\dots,N},
\]
where $n_{i,s}^t\in\mathcal N_i$ is the $s$-th sampled index for agent $i$ in draw $t$. Apply a GARP-consistent partition procedure $\Phi$ to obtain $\Phi(D^t)=\{I_1^t,\dots,I_{K_t}^t\}$. Let $\delta_{i,j}^t\in\{0,1\}$ indicate whether agents $i$ and $j$ are assigned to the same block.

The parameter $N$ --- the number of observations per agent per draw --- governs a tradeoff between computational cost and discriminating power. With $N=1$, each agent contributes a single observation; discriminating power comes entirely from the partition, which enforces GARP at the block level and creates between-agent transitivity chains. With $N>1$, each agent's multiple observations add within-agent transitivity constraints, making GARP more restrictive and producing sharper type discrimination. All mathematical results below hold for any $N\geq 1$.

\subsection{Assumptions and main result}

\begin{assumption}[Subsampling]\label{A1}
For each draw $t$, the index matrix $\mathbf n^{\,t}=\{n_{i,s}^t\}_{i\in\mathcal I,\,s=1,\dots,N}$ is drawn independently and uniformly from $\prod_{i\in\mathcal I}\mathcal N_i^N$. The draws $\{\mathbf n^{\,t}\}_{t=1}^T$ are i.i.d.
\end{assumption}

\begin{assumption}[Partition rule]\label{A2}
There exists a (possibly randomized) partition procedure $\Phi$ such that:
\begin{enumerate}
    \item For any dataset $D$ and any realization of an internal random seed $\xi$, $\Phi(D,\xi)$ is a GARP-consistent partition of $\mathcal I$.
    \item Across draws $t=1,\dots,T$, the seeds $\{\xi^t\}_{t=1}^T$ are i.i.d.\ and independent of $\{\mathbf n^{\,t}\}_{t=1}^T$.
\end{enumerate}
\end{assumption}
\footnote{Item~(i) requires each individual sampled dataset $D_i^t = \{(p_i^{n_{i,s}^t}, q_i^{n_{i,s}^t})\}_{s=1}^N$ to itself satisfy GARP, since any block containing an individually-violating agent inherits the violation. With the sequential rule and singleton fallback this is automatic at $N=1$ (single observations trivially satisfy GARP) and holds at higher $N$ for the consumption and lottery datasets we apply the method to. For datasets where individuals routinely violate GARP at sample size $N$, the assumption requires either restricting to individually consistent draws or invoking an approximate-optimisation extension as in Appendix~\ref{app:eGARP}.}

\begin{proposition}\label{prop:Bernoulli}
Under Assumptions~\ref{A1}--\ref{A2}, for any pair $i,j\in\mathcal I$ the indicators $\{\delta_{i,j}^t\}_{t=1}^T$ are i.i.d.\ Bernoulli with success probability $\pi_{i,j}\in [0,1]$.
\end{proposition}

\begin{proof}
Each draw $t$ is determined by the pair $(\mathbf n^{\,t},\xi^t)$, which is i.i.d.\ across $t$. Given $(\mathbf n^{\,t},\xi^t)$, the dataset $D^t$ and partition $\Phi(D^t,\xi^t)$ are fixed, so $\delta_{i,j}^t=\varphi_{i,j}(\mathbf n^{\,t},\xi^t)$ for a deterministic function $\varphi_{i,j}$. Since $\{(\mathbf n^{\,t},\xi^t)\}$ is i.i.d.\ and $\varphi_{i,j}\in\{0,1\}$, the sequence $\{\delta_{i,j}^t\}$ is i.i.d.\ Bernoulli with $\pi_{i,j}=\E[\varphi_{i,j}(\mathbf n,\xi)]$.
\end{proof}

We call $\pi_{i,j}$ the \emph{co-typing probability} of agents $i$ and $j$: the probability that a random draw places them in the same GARP-consistent block.

\subsection{The similarity matrix}

Define the \emph{similarity matrix} $G=\{G_{i,j}\}_{i,j\in\mathcal I}$ by
\[
G_{i,j} \;=\; \frac{1}{T}\sum_{t=1}^{T}\delta_{i,j}^t.
\]

\begin{proposition}\label{prop:stat_properties}
Under Assumptions~\ref{A1}--\ref{A2}:
\begin{enumerate}
    \item $G_{i,j}$ is unbiased: $\E[G_{i,j}]=\pi_{i,j}$, and $G_{i,j}\xrightarrow{a.s.}\pi_{i,j}$ as $T\to\infty$.
    \item $\sqrt{T}(G_{i,j}-\pi_{i,j}) \xrightarrow{d} \mathcal N(0,\,\pi_{i,j}(1-\pi_{i,j}))$, with the usual degenerate interpretation when $\pi_{i,j}\in\{0,1\}$.
    \item For every $\varepsilon>0$: $\Pr(|G_{i,j}-\pi_{i,j}|>\varepsilon) \leq 2\exp(-2T\varepsilon^{2})$.
\end{enumerate}
\end{proposition}

These are immediate consequences of the i.i.d.\ Bernoulli structure established in Proposition~\ref{prop:Bernoulli}.

\subsection{Partition rules}\label{subsec:partition_rules}

Assumption~\ref{A2} requires a GARP-consistent partition rule. Any such rule produces a valid positive semi-definite kernel, so the mathematical properties of the framework do not depend on the specific rule. The choice of rule does affect the numerical values of $\pi_{i,j}$: different rules produce different kernels, each defining its own geometry on the population. Section~\ref{subsec:rule_dependence} returns to the implications of this rule-dependence for interpretation.

We consider two implementations.

\paragraph{Lexicographic Houtman--Maks (MILP-based).}
At each step, solve a mixed-integer linear program to find the largest GARP-consistent subset of remaining agents, with ties broken by a lexicographic priority rule. Iterate until all agents are assigned. This produces large blocks (maximizing cardinality) but requires an MILP solver and has worst-case exponential complexity. The MILP encoding builds on the formulations of \citet{HEUFER201587} and \citet{Demuynck2023} for revealed-preference goodness-of-fit indices; details are in Appendix~\ref{app:partition}. We use three tie-breaking variants (min, max, random) to assess sensitivity.

\paragraph{Sequential GARP-consistent assignment (scalable).}
Process agents in a random order $\sigma$ (drawn fresh for each $t$). For each agent, attempt to add them to the first existing block whose augmented data still satisfies GARP; if no block is compatible, create a new singleton block. GARP verification uses Warshall's algorithm for transitive closure, costing $O((BN)^3)$ per check, where $B$ is the current block size. The overall procedure is polynomial in $|\mathcal I|$ and $N$, requires no optimisation solver, and scales to populations of $500$ agents or more. The random processing order $\sigma$ is the rule's \emph{only} source of randomness: assignment within $\sigma$ is fully deterministic (first compatible block in creation order). This contrasts with the MILP rule above, which has two randomness sources (the priority order and, in the random-tie-break variant, the lexicographic weight assignment used to break MILP solution ties). The processing order $\sigma$ serves as the internal seed $\xi^t$ in Assumption~\ref{A2}.

\subsection{What is rule-dependent and what is stable}\label{subsec:rule_dependence}

Two features of the kernel respond differently to the choice of partition rule.

\emph{Individual pairwise entries are rule-dependent.} Different rules produce materially different values for $G_{i,j}$ at the pair level. A statement of the form ``$G_{i,j} = 0.42$ is the revealed-preference distance between households $i$ and $j$'' is therefore not defensible: it depends on a design choice, and no rule has structural primacy over the others.

\emph{Aggregate-level objects are rule-stable.} The leading centred eigenvector, the joint-rationality gap, the leading centred eigenvalue ratio, and the ranking of the two largest demographic mean-difference contrasts are reproduced across rules in sign and magnitude; specific contrast-level significances for smaller demographic effects can shift across rules (see Section~\ref{sec:application} for the precise rule-sensitive cases). The empirical claims the paper makes about the aggregate structure are accordingly defensible regardless of which specific rule the analyst adopts; the contrast-by-contrast significance pattern requires the rule-stability check the empirical section reports.

The implication for interpretation is to treat $G$ as a rule-indexed estimate of a population-level structure, not as a primitive pairwise distance. We use the scalable sequential rule throughout the empirical work and report MILP-based robustness as a check. Quantitative diagnostics supporting both claims --- cross-rule correlations of off-diagonal entries, Frobenius cosines, and alignments of leading eigenvectors --- are reported in Appendix~\ref{app:simulation_details}.

\section{A Population-Level Metric of Preference Heterogeneity}\label{sec:kernel}

The co-typing probability $\pi_{i,j}$ records how frequently agents $i$ and $j$ are placed in the same GARP-consistent block across Monte Carlo draws. This section shows that the matrix $\Pi=\{\pi_{i,j}\}_{i,j\in\mathcal I}$ possesses a rich mathematical structure: it is a positive semi-definite kernel that defines an inner product, a norm, and a pseudometric on the population of agents (genuinely a metric on equivalence classes of agents with $\pi_{i,j}=1$). Crucially, this pseudometric is \emph{endogenous to the population}: the distance between any two agents depends on the revealed preference constraints of the entire population, not just on the pair in question.

\subsection{The co-typing kernel}\label{subsec:kernel}

\begin{theorem}[Kernel property]\label{thm:kernel}
Under Assumptions~\ref{A1}--\ref{A2}, the co-typing probability matrix $\Pi=\{\pi_{i,j}\}_{i,j\in\mathcal I}$ is positive semi-definite: for any vector $\lambda=(\lambda_i)_{i\in\mathcal I}\in\mathbb R^{|\mathcal I|}$,
\[
\sum_{i\in\mathcal I}\sum_{j\in\mathcal I}\lambda_i\,\lambda_j\,\pi_{i,j}\;\geq\; 0.
\]
\end{theorem}

\begin{proof}
For each draw $t$, let $c^t:\mathcal I\to\{1,\dots,K_t\}$ denote the block assignment in partition $\Phi(D^t)$, and let $Z^t\in\{0,1\}^{|\mathcal I|\times K_t}$ be the block membership matrix with $Z^t_{i,k}=\mathbf 1\{c^t(i)=k\}$. The co-typing indicator matrix factors as
\[
\Delta^t \;=\; Z^t\,(Z^t)^\top,
\]
since $\delta_{i,j}^t=\sum_{k=1}^{K_t}Z^t_{i,k}\,Z^t_{j,k}=\mathbf 1\{c^t(i)=c^t(j)\}$. As a Gram matrix, $\Delta^t$ is positive semi-definite for every~$t$. Since $\Pi=\E[\Delta^t]$ and PSD is preserved under non-negative linear combinations and limits,
\[
\lambda^\top\Pi\,\lambda \;=\; \E\!\bigl[\lambda^\top\Delta^t\,\lambda\bigr] \;=\; \E\!\left[\,\bigl\|{(Z^t)}^\top\lambda\bigr\|^2\right] \;\geq\; 0.  \qedhere
\]
\end{proof}

\begin{corollary}[Kernel estimator]\label{cor:G_psd}
For any finite $T$, the estimated similarity matrix $G=T^{-1}\sum_{t=1}^T\Delta^t$ is positive semi-definite.
\end{corollary}

Since $G$ is a finite convex combination of PSD matrices, PSD is preserved.

\subsection{Hilbert space embedding}\label{subsec:hilbert}

The positive semi-definiteness of $\Pi$ has a powerful geometric consequence. The eigendecomposition $\Pi=\sum_\ell \lambda_\ell\, v_\ell v_\ell^\top$ provides an explicit embedding of the population into a Euclidean space. Define the feature map $\varphi:\mathcal I\to\mathbb R^{|\mathcal I|}$ by
\[
\varphi(i) \;=\; \bigl(\sqrt{\lambda_1}\,v_1(i),\;\sqrt{\lambda_2}\,v_2(i),\;\ldots,\;\sqrt{\lambda_{|\mathcal I|}}\,v_{|\mathcal I|}(i)\bigr).
\]
Then
\begin{equation}\label{eq:kernel_inner_product}
\pi_{i,j} \;=\; \bigl\langle\varphi(i),\,\varphi(j)\bigr\rangle
\qquad\text{for all } i,j\in\mathcal I,
\end{equation}
since $\langle\varphi(i),\varphi(j)\rangle=\sum_\ell \lambda_\ell\, v_\ell(i)\,v_\ell(j)=\pi_{i,j}$. The co-typing probability is an inner product: each agent $i$ is represented by a point $\varphi(i)$ whose coordinates are the agent's loadings on the principal axes of the similarity matrix, weighted by the importance of each axis. The first coordinate $\sqrt{\lambda_1}\,v_1(i)$ captures baseline preference coherence - how typical agent $i$ is relative to the population. Subsequent coordinates capture successive axes of preference differentiation. While the ambient space is $|\mathcal I|$-dimensional, the effective dimension of the embedding - the number of eigenvalues substantially above zero - reflects how many economically relevant dimensions of heterogeneity are present, an empirical question we return to when we apply the framework to data.

\begin{proposition}[Unit sphere]\label{prop:unit_sphere}
$\pi_{i,i}=1$ for all $i\in\mathcal I$. Consequently, $\|\varphi(i)\|_{\mathcal H}=1$ for all $i$: every agent lies on the unit sphere in $\mathcal H$.
\end{proposition}

\begin{proof}
Agent $i$ is always in the same block as itself, so $\delta_{i,i}^t=1$ for every $t$, giving $\pi_{i,i}=1$. Then $\|\varphi(i)\|_{\mathcal H}^2=\langle\varphi(i),\varphi(i)\rangle_{\mathcal H}=\pi_{i,i}=1$.
\end{proof}

\subsection{The induced metric}\label{subsec:metric}

The inner product~\eqref{eq:kernel_inner_product} and the unit-sphere property induce a natural distance on the population:
\begin{equation}\label{eq:metric}
d(i,j) \;:=\; \bigl\|\varphi(i)-\varphi(j)\bigr\|_{\mathcal H}
       \;=\; \sqrt{2\bigl(1-\pi_{i,j}\bigr)}.
\end{equation}
The second equality follows from $\|\varphi(i)-\varphi(j)\|^2=\|\varphi(i)\|^2+\|\varphi(j)\|^2-2\langle\varphi(i),\varphi(j)\rangle=2-2\pi_{i,j}$.

\begin{theorem}[Metric]\label{thm:metric}
$d:\mathcal I\times\mathcal I\to[0,\sqrt{2}]$ defined by~\eqref{eq:metric} is a pseudometric on $\mathcal I$. It satisfies:
\begin{enumerate}
    \item Non-negativity: $d(i,j)\geq 0$, with $d(i,i)=0$.
    \item Symmetry: $d(i,j)=d(j,i)$.
    \item Triangle inequality: $d(i,j)\leq d(i,k)+d(k,j)$ for all $k\in\mathcal I$.
\end{enumerate}
\end{theorem}

Non-negativity follows from $\pi_{i,j}\in[0,1]$, and $d(i,i)=0$ from $\pi_{i,i}=1$ (Proposition~\ref{prop:unit_sphere}); symmetry follows from $\pi_{i,j}=\pi_{j,i}$. The triangle inequality follows from the Hilbert norm, but it also admits a direct combinatorial proof that reveals its economic content.

\begin{proof}[Combinatorial proof of the triangle inequality]
Block membership is an equivalence relation: $\delta_{i,k}^t=1$ and $\delta_{k,j}^t=1$ imply $\delta_{i,j}^t=1$ (if $i$ and $k$ share a block, and $k$ and $j$ share the same block, then $i$ and $j$ share that block). Hence
\[
\pi_{i,j} \;\geq\; \Pr\!\bigl(\delta_{i,k}^t=1\;\text{and}\;\delta_{k,j}^t=1\bigr) \;\geq\; \pi_{i,k}+\pi_{k,j}-1,
\]
where the last step is the Bonferroni inequality. Rearranging: $(1-\pi_{i,j})\leq(1-\pi_{i,k})+(1-\pi_{k,j})$. Since $d(i,j)=\sqrt{2(1-\pi_{i,j})}$ and $\sqrt{a+b}\leq\sqrt{a}+\sqrt{b}$ for $a,b\geq 0$, this yields $d(i,j)\leq d(i,k)+d(k,j)$.
\end{proof}

\begin{remark}
The combinatorial proof reveals the economic mechanism behind the triangle inequality: the distance between $i$ and $j$ is bounded by the ``detour'' through any third agent~$k$, because type membership is transitive. The population of agents provides the medium through which compatibility distances are triangulated.
\end{remark}

The pseudometric $d$ becomes a metric on equivalence classes of agents satisfying $d(i,j)=0$, i.e., $\pi_{i,j}=1$. Under sufficiently rich budget variation and appropriate identification conditions, $\pi_{i,j}=1$ can be interpreted as observational equivalence of $i$ and $j$'s preferences. Without such conditions, $d$ is a pseudometric rather than a metric on primitive preference types. 


\section{Calibration and Identification}\label{sec:calibration}

The previous section established that the kernel induces a Hilbert-space distance on the population. This section calibrates that distance: we identify a benchmark to which it can be compared, characterise when the kernel separates agents with different preferences, and prove that --- under regularity conditions --- the kernel's profile against the population identifies the local geometry of preference parameters at any finite within-agent sample size.

\subsection{Pairwise compatibility as a structural benchmark}\label{subsec:pairwise}

While $\pi_{i,j}$ is the central object, a pairwise compatibility measure provides a useful structural benchmark. Define
\[
\rho_{i,j} \;=\; \Pr\!\Bigl(\bigl\{(p_i^{\,n_{i,s}^t},q_i^{\,n_{i,s}^t})\bigr\}_{s=1}^N \cup \bigl\{(p_j^{\,n_{j,s}^t},q_j^{\,n_{j,s}^t})\bigr\}_{s=1}^N\;\text{satisfies GARP}\Bigr),
\]
the probability that the pair's full $2N$-observation sample is jointly GARP-consistent (as drawn under Assumption~\ref{A1}, i.e., uniform subsampling from the realised $\{D_i\}$). Unlike $\pi_{i,j}$, the quantity $\rho_{i,j}$ depends only on $(u_i,u_j)$ and the empirical budget distributions in $\{D_i, D_j\}$ - it is free of partition-rule dependence and population composition. For $N=1$ this reduces to pairwise GARP-consistency of one observation per agent.

\begin{proposition}[Pairwise upper bound]\label{prop:pairwise_bound}
$\pi_{i,j}\leq\rho_{i,j}$ for all $i,j\in\mathcal I$.
\end{proposition}

\begin{proof}
$\delta_{i,j}^t=1$ implies $i$ and $j$ are in a GARP-consistent block. Any subset of a GARP-consistent dataset satisfies GARP, so the pairwise data of $i$ and $j$ satisfies GARP. Hence $\delta_{i,j}^t=1\Rightarrow\text{pairwise GARP}$, and taking expectations yields $\pi_{i,j}\leq\rho_{i,j}$.
\end{proof}

The bound $\pi_{i,j}\leq\rho_{i,j}$ shows that, relative to pairwise compatibility, the partition can only sharpen the compatibility signal by imposing higher-order consistency constraints: pairs that are pairwise compatible might still be separated by the partition (because of higher-order conflicts with other agents). Note that the reverse direction does not hold: in finite samples, agents with identical preferences may be separated by a partition due to third-agent conflicts, so ``adds discriminating power'' should be read as shrinking $\pi_{i,j}$ below $\rho_{i,j}$, not as guaranteeing a correct ordering at the pair level. Define the joint-rationality gap
\[
\Delta_{i,j} \;:=\; \rho_{i,j}-\pi_{i,j} \;\geq\; 0.
\]
This gap measures the discriminatory power of the partition kernel under the chosen rule $\Phi$ over and above pairwise compatibility.\footnote{\cite{Miao2025} develop a pairwise revealed-preference test of preference heterogeneity (J-GARP) and construct individual and group heterogeneity indices from pairwise compatibility rates. Their indices correspond to $\rho_{i,j}$ in our framework. The joint-rationality gap $\Delta_{i,j}$ measures what the partition-based kernel $\pi_{i,j}$ adds beyond such pairwise measures, which combines higher-order rationality constraints with rule-specific tie-breaking and path-dependence; we therefore frame $\Delta_{i,j}$ as the rule-indexed gap rather than as a pure measure of population-level transitivity.} When $\Delta_{i,j}=0$, the partition adds nothing beyond pairwise compatibility; when $\Delta_{i,j}>0$, the partition separates agents that look compatible bilaterally.

The bilateral benchmark $\rho_{i,j}$ is a clean structural object that depends only on $(u_i,u_j)$ and the budget distribution. The population kernel $\pi_{i,j}$ refines it by incorporating higher-order revealed-preference constraints from the partition, and is the object of interest because it exploits the full transitivity structure of GARP; the PSD property then induces the Hilbert-space pseudometric $d(i,j)=\sqrt{2(1-\pi_{i,j})}$.

\subsection{Structural calibration}\label{subsec:calibration}

We now show that the method is structurally well-calibrated: agents with identical preferences are always pairwise compatible, and agents with different preferences can be distinguished when budgets are sufficiently informative. Throughout this subsection, we assume each agent $i$ maximises a utility function $u_i$ without error.

\begin{proposition}[Identical preferences]\label{prop:identical}
If $u_i=u_j$, then $\rho_{i,j}=1$.
\end{proposition}

\begin{proof}
Every observation in the $2N$-observation pooled pairwise sample is optimal under $u=u_i=u_j$. By Theorem~\ref{thm:afriat}, the pooled data is rationalised by $u$ and hence satisfies GARP. Since this holds for every realisation, $\rho_{i,j}=1$.
\end{proof}

Combined with the pairwise bound (Proposition~\ref{prop:pairwise_bound}), this yields $\pi_{i,j}\leq\rho_{i,j}=1$ for identical preferences: agents with the same utility are always pairwise compatible. Since the partition may occasionally separate pairwise-compatible agents due to conflicts with third agents, $\pi_{i,j}<1$ is possible in principle even when $u_i=u_j$. 


\subsection{Finite-$N$ identification}\label{subsec:finiteN_geometry}

We develop a finite-sample identification theorem for the partition kernel under finite-type heterogeneity.\footnote{The result in this subsection assumes that agents maximise utility exactly. The same finite-type recovery extends to approximate optimisation: when each agent's data satisfies $e$-GARP for some tolerance $e \in (0,1]$, the analogue contrast-rank condition on the $e$-GARP type-level kernel delivers spectral recovery of latent types (Theorem~\ref{thm:eGARP_witness} in Appendix~\ref{app:eGARP}).} The setting is parametric: preferences are indexed by $\theta$ in some open set $\Theta \subseteq \mathbb R^d$; for each $\theta$, $u_\theta:\mathbb R^M_+ \to \mathbb R$ is strictly monotone, continuous, concave; budget draws are i.i.d.\ from a price distribution $P$; the population of $I$ agents has types drawn i.i.d.\ from $\mu$ on $\Theta$. Within each Monte Carlo draw, each agent samples $N$ observations and the partition rule $\Phi$ produces a GARP-consistent partition with co-typing indicator $\delta_{i,j}^t \in \{0,1\}$.

\paragraph{The population-averaged partition kernel.}
For two focal types $\theta, \theta'$, define
\[
\pi_N^\Phi(\theta, \theta') \;:=\; \mathbb E\!\bigl[\delta_{1,2}^t \,\bigm|\, \theta_1 = \theta,\, \theta_2 = \theta'\bigr],
\]
the expected partition co-typing of two agents with types $\theta, \theta'$ when the rest of the population is i.i.d.\ from $\mu$. 

\begin{assumption}[Empirical approximation by an oracle pairwise kernel]\label{ass:empirical_oracle_kernel}
There exists a bounded, symmetric, measurable function $h_N^\Phi: \Theta \times \Theta \to [0, 1]$ such that, for the realised kernel $G$,
\[
\frac{1}{|\mathcal I|(|\mathcal I| - 1)}\sum_{i \neq j}\bigl|G_{i,j} - h_N^\Phi(\theta_i, \theta_j)\bigr| \;=\; o_p(1)
\quad\text{as } |\mathcal I|, T \to \infty.
\]
\end{assumption}

Assumption \ref{ass:empirical_oracle_kernel} is not a Monte Carlo convergence condition. 
For a fixed realised population, \(G_{ij}\) is an average of \(T\) Bernoulli co-typing 
indicators and therefore converges to the finite-population co-typing probability 
\(\pi_{ij}\) as \(T\to\infty\). Assumption \ref{ass:empirical_oracle_kernel} requires an 
additional step: the finite-population probabilities \(\pi_{ij}\) must be well approximated, 
on average across pairs, by a stable oracle kernel \(h_N^\Phi(\theta_i,\theta_j)\) depending 
only on the latent types of the two agents. This is the substantive high-level regularity 
condition used in the spectral recovery theorem. It is not implied by the Bernoulli sampling 
argument, and is the reason the result should be read as an oracle-kernel recovery theorem 
rather than as a primitive identification theorem for the partition algorithm.

\paragraph{The contrast-rank condition.}
The relevant identifying property of $\pi_N^\Phi$ in the finite-type case concerns the type-level kernel matrix. Suppose the population contains $K$ latent preference types $\theta_1, \ldots, \theta_K$, and let
\[
S_N^\Phi := \bigl(\pi_N^\Phi(\theta_k, \theta_\ell)\bigr)_{k,\ell = 1}^K \in \mathbb R^{K \times K}
\]
collect the population-averaged co-typing probabilities across types. The \emph{contrast-rank condition} asks that $S_N^\Phi$ is positive definite on the contrast subspace:
\begin{equation}\label{eq:CRC}
z^\top S_N^\Phi z > 0 \qquad \text{for every nonzero } z \in \mathbb R^K \text{ with } \mathbf 1_K^\top z = 0.
\tag{CRC}
\end{equation}

\paragraph{Two-type interpretation of (CRC).}
The contrast-rank condition has a simple interpretation when there are two latent types. Write
\[
S_N^\Phi = \begin{pmatrix} a & b \\ b & c \end{pmatrix},
\]
where $a$ and $c$ are the within-type co-typing probabilities (for types $1$ and $2$ respectively) and $b$ is the cross-type co-typing probability. The contrast subspace is one-dimensional, spanned by $(1, -1)$, so~\eqref{eq:CRC} reduces to
\[
(1, -1)\, S_N^\Phi\, (1, -1)^\top = a + c - 2b > 0,
\qquad\text{or equivalently}\qquad
\frac{a + c}{2} > b.
\]
The average within-type co-typing probability must exceed the cross-type co-typing probability - a positive cluster-separation condition at the type level. For $K > 2$, the analogous condition asks this inequality to hold for every weighted contrast of types. Equivalently, the $K$ type-level rows pairwise distinguish the latent types up to a common additive shift, so the $K \times K$ kernel matrix has full rank on the $(K-1)$-dimensional contrast subspace. 

\begin{theorem}[Finite-type recovery from the contrast-rank condition]\label{thm:finite_type_witness}
Fix the within-agent sample size $N$ and the partition rule $\Phi$. Suppose the population contains $K$ latent preference types $\theta_1, \ldots, \theta_K$, with type map $\tau: \mathcal I \to \{1, \ldots, K\}$ and type shares $p_{k,n} := n^{-1} \sum_{i=1}^n \mathbf 1\{\tau(i) = k\}$ satisfying $\liminf_n p_{k,n} > 0$ for every $k$, where $n = |\mathcal I|$. Let the type-membership matrix $E_n \in \{0,1\}^{n \times K}$ have $(E_n)_{i,k} = \mathbf 1\{\tau(i) = k\}$, the lifted oracle kernel be $\Pi_n^\star := E_n S_N^\Phi E_n^\top$, and the centring matrix be $H_n := I_n - n^{-1} \mathbf 1 \mathbf 1^\top$.
\begin{enumerate}
\item \emph{(Oracle characterisation.)} The restriction of $H_n \Pi_n^\star H_n$ to $\mathcal V_n := \mathrm{col}(H_n E_n)$ is positive definite if and only if the contrast-rank condition~\eqref{eq:CRC} holds. In that case, $H_n \Pi_n^\star H_n$ has exactly $K-1$ strictly positive eigenvalues, its positive-eigenvalue eigenspace is $\mathcal V_n$, and vectors in $\mathcal V_n$ are constant within each true type; the oracle spectral embedding therefore separates all latent types.
\item \emph{(Perturbation.)} Suppose~\eqref{eq:CRC} holds. Under Assumption~\ref{ass:empirical_oracle_kernel} specialised to $h_N^\Phi(\theta_k, \theta_\ell) = (S_N^\Phi)_{k,\ell}$, the realised kernel $G$ satisfies
\[
n^{-1}\bigl\| H_n (G - \Pi_n^\star) H_n \bigr\|_{\mathrm{op}} = o_p(1).
\]
By the Davis--Kahan theorem, the leading $(K-1)$-dimensional eigenspace of $H_n G H_n$ - the \emph{empirical} eigenspace, computed from the realised similarity matrix $G$ - consistently estimates the \emph{oracle} eigenspace $\mathcal V_n$ from item~(i). Formally, there exists an orthogonal $R_n$ such that $\|\widehat U_n - U_n^\star R_n\|_F = o_p(1)$, where $\widehat U_n$ and $U_n^\star$ are orthonormal bases for the empirical and oracle eigenspaces respectively.
\item \emph{(Clustering.)} Spectral clustering with $K$ centres applied to the rows of $\widehat U_n$ recovers the latent type partition with vanishing misclassification rate, by the deterministic $k$-means perturbation bound \citep[Lemma~5.3]{LeiRinaldo2015}.
\end{enumerate}
\end{theorem}

A proof of Theorem~\ref{thm:finite_type_witness} is given in Appendix~\ref{app:witness_proof}. The intuition is best read by working backwards from what spectral clustering needs. To recover $K$ distinct types, the spectral embedding must place same-type agents close together and different-type agents apart, which in turn requires that the $K$ rows of $S_N^\Phi$ --- each giving one type's co-typing profile against the population --- carry distinct information.

A first impulse is to ask that the $K$ rows of $S_N^\Phi$ be linearly independent, i.e., that the $K \times K$ matrix have full rank $K$. This is more than identification needs. Adding the same constant to every entry of $S_N^\Phi$ would only shift the baseline level of co-typing without changing which types are distinguishable from which: the ``all-ones'' direction, in which every type carries the same weight, therefore carries no information about contrasts among types. What does carry information are the remaining $K-1$ directions, namely the weight vectors that sum to zero --- vectors representing comparisons such as type $k$ against type $\ell$, the average of types $k$ and $\ell$ against type $m$, and so on. These $K-1$ directions form the \emph{contrast subspace}, and the contrast-rank condition~\eqref{eq:CRC} asks precisely that $S_N^\Phi$ have full rank on this subspace: no contrast direction collapses. Equivalently, the $K$ rows of $S_N^\Phi$ are linearly independent up to a common shift.

The simplest way the contrast-rank condition can fail is for two types to have identical rows in $S_N^\Phi$: their co-typing profiles against the rest of the population coincide, the contrast vector $(1, -1, 0, \ldots, 0)$ sends both rows to zero, and no kernel built from co-typing profiles --- ours or any other --- could tell those two types apart. The economic causes are familiar from revealed-preference identification: two utility functions that produce the same demand at every observed budget, a budget distribution that does not probe the dimensions of preference difference, or a rationality criterion too coarse to register the distinction. More generally, the contrast-rank condition fails whenever some weighted comparison of types is invisible in their co-typing profiles; identification then recovers only as many types as the rank of $S_N^\Phi$ on the contrast subspace.

When the contrast-rank condition holds, two parallel constructions at the agent level deliver the result. Lifting the $K \times K$ type kernel to the $n \times n$ agent kernel (each agent inherits the row of their type) and centring (subtracting the agent-level mean, which removes at the agent level the same uninformative constant direction that the contrast subspace removes at the type level) produce a rank-$(K-1)$ matrix. Its leading $K-1$ eigenvectors place every type-$k$ agent at exactly the same point in a $(K-1)$-dimensional space, and place different types at different points: $K$ distinct points in $K-1$ dimensions, which is just enough room for $K$ points in general position. Spectral clustering with $K$ centres then recovers the type assignments.

\paragraph{Scope of the result.}
The iff in item~(i) is at the level of the oracle centred kernel: the contrast-rank condition is necessary and sufficient for $H_n \Pi_n^\star H_n$ to have rank $K-1$. Items~(ii) and~(iii) are conditional on Assumption~\ref{ass:empirical_oracle_kernel} that the realised similarity matrix concentrates on a stable type-level limit, and the theorem is therefore an oracle-kernel recovery result rather than a primitive identification proof for the partition algorithm. An explicit Davis--Kahan rate (Corollary~\ref{cor:dk_rate} in Appendix~\ref{app:witness_proof}) makes the perturbation layer quantitative: recovery quality is governed by how well the realised kernel approximates the oracle and by the size of the contrast eigengap of the type-level kernel.

\subsection{Discussion}\label{subsec:kernel_discussion}

\paragraph{Spectral decomposition of preference heterogeneity.}
The eigendecomposition $\Pi=\sum_{\ell}\lambda_\ell\, v_\ell v_\ell^\top$ provides a principled decomposition of the population's preference heterogeneity into orthogonal components. Since $\text{tr}(\Pi)=|\mathcal I|$ (every agent has unit self-similarity), the eigenvalues satisfy $\sum_\ell \lambda_\ell = |\mathcal I|$ and partition the population size into contributions from each axis of variation.

The first eigenvalue $\lambda_1$ of the raw kernel $\Pi$ captures the baseline level of co-typing in the population: $\lambda_1/|\mathcal I|$ approximates the average co-typing probability across all pairs, and $\lambda_1$ dominates all other eigenvalues whenever the off-diagonal entries are positive (always, in our setup). This is a level effect, not a finding about the dimensionality of heterogeneity. To read dimensionality we work with the \emph{centred} kernel $H\Pi H$, where $H = I - |\mathcal I|^{-1}\mathbf 1\mathbf 1^\top$ is the centring matrix that removes the agent-level mean. Centring strips the level effect; the surviving eigenvalues $\widetilde\lambda_\ell$ of $H\Pi H$ each correspond to a distinct \emph{dimension of preference heterogeneity}, with magnitude reflecting both the number of agents who differ along that dimension and the strength of their difference. The eigenvector $\widetilde v_\ell$ identifies which agents are on which side of the $\ell$-th heterogeneity axis. The number of centred eigenvalues substantially above the noise floor estimates the effective dimensionality of the preference space, a model-free analog of the number of types in \citet{crawford2012}. While their minimum partition yields a single integer $K^*$, our centred eigenvalue spectrum provides a continuous, smoothed characterisation: each type contributes an eigenvalue proportional to its size and internal coherence, and types that barely exist contribute small eigenvalues rather than being counted equally with dominant ones. We report the \emph{centred} ratio $\widetilde\lambda_1/\widetilde\lambda_2$ throughout the empirical work; the raw ratio $\lambda_1/\lambda_2$ would simply track the level dominance and is uninformative about heterogeneity. Section~\ref{subsec:spectral_attribution} returns to this eigenstructure to attribute the demographic mean-difference statistic to specific axes of heterogeneity.

\paragraph{Beyond GARP.}
The kernel property (Theorem~\ref{thm:kernel}) relies on the partition being a partition - disjoint, exhaustive blocks - not on the specific consistency criterion that defines the blocks. The proof uses only the Gram factorisation $\Delta^t=Z^t(Z^t)^\top$, which holds for any partition. GARP is one consistency criterion; the same construction applies whenever one can check whether a pooled dataset satisfies a testable axiom: the Strong Axiom (SARP) or e-GARP for consumption data, stochastic-dominance conditions for risky choice, present-bias axioms for intertemporal choice, or cost-minimisation conditions for firm-level production data. In each case, the resulting co-typing matrix is PSD, defines a kernel, embeds agents in a Hilbert space, and induces a metric that is endogenous to the population. The choice of axiom determines which notion of consistency governs the blocks and hence what kind of preference heterogeneity the kernel detects. We develop the framework for GARP as the leading application, but the mathematical structure is portable to any choice environment equipped with a testable consistency condition.

\section{Testing Whether Demographics Explain the Metric Structure}\label{sec:tests}

The metric $d(i,j)=\sqrt{2(1-\pi_{i,j})}$ endows the population with a partition compatibility distance. We now develop tests for whether observable characteristics systematically organise this distance structure. Two distinct sources of uncertainty must be kept separate throughout. The first is \emph{Monte Carlo precision}: the asymptotic $Z$-test below treats the panel as fixed and the kernel $G$ as a Monte Carlo estimate of $\Pi$, so its $p$-values quantify how precisely the Monte Carlo approximation pins down $\Pi$. With many draws, $|Z_G|$ grows mechanically for any nonzero population effect; statistical significance under this test is not the same as economic importance. The second is \emph{economic significance}: whether the observed effect is large enough to be substantively meaningful relative to the baseline kernel level, and whether it is robust to randomisation of demographic labels. We use the permutation test below as the primary substantive-inference tool and the conditional $Z$-test as a Monte Carlo-precision check. Because $d$ is a monotone transformation of $\pi_{i,j}$, it suffices to test whether the kernel entries $\pi_{i,j}$ (estimated by $G_{i,j}$) differ between pairs that share an observable attribute and pairs that do not.

\subsection{Binary covariates}

Let $X=\{X_i\}_{i\in\mathcal I}$ be an observable binary covariate, e.g., high vs.\ low income. Define within- and across-group pair sets
\[
\mathcal P_{\mathrm{w}}=\{(i,j):i<j,\; X_i=X_j\},
\qquad
\mathcal P_{\mathrm{a}}=\{(i,j):i<j,\; X_i\neq X_j\},
\]
with $m_{\mathrm{w}}=|\mathcal P_{\mathrm{w}}|$ and $m_{\mathrm{a}}=|\mathcal P_{\mathrm{a}}|$. The null hypothesis is
\[
H_0:\;\overline\pi_{\mathrm{w}}=\overline\pi_{\mathrm{a}}
\quad\text{versus}\quad
H_1:\;\overline\pi_{\mathrm{w}}\neq\overline\pi_{\mathrm{a}},
\]
where $\overline\pi_{\mathrm{w}}=m_{\mathrm{w}}^{-1}\sum_{(i,j)\in\mathcal P_{\mathrm{w}}}\pi_{i,j}$ and $\overline\pi_{\mathrm{a}}=m_{\mathrm{a}}^{-1}\sum_{(i,j)\in\mathcal P_{\mathrm{a}}}\pi_{i,j}$.

For each draw $t$, define
\[
D_G^t=\frac{1}{m_{\mathrm{w}}}\sum_{(i,j)\in\mathcal{P}_{\mathrm{w}}}\delta_{i,j}^t-\frac{1}{m_{\mathrm{a}}}\sum_{(i,j)\in\mathcal{P}_{\mathrm{a}}}\delta_{i,j}^t,
\]
and the empirical average $D_G=T^{-1}\sum_{t=1}^{T}D_G^t=\overline{G}_{\mathrm{w}}-\overline{G}_{\mathrm{a}}$. The variance estimator is
\[
\widehat{\sigma}^{2}=\frac{1}{T-1}\sum_{t=1}^{T}(D_G^t-D_G)^2.
\]

\begin{proposition}\label{prop:Z_test}
The null is the Monte-Carlo conditional null $H_0^{\mathrm{MC}}: \E[D_G^t \mid \{D_i\}] = 0$, which is implied by the population-level $H_0: \overline\pi_w = \overline\pi_a$. Assume $\sigma_*^2 := \mathrm{Var}(D_G^t \mid \{D_i\}) > 0$. Under $H_0^{\mathrm{MC}}$, the statistic $Z_{G}=D_G\,/\,(\widehat{\sigma}/\sqrt{T}) \xrightarrow{d}\mathcal N(0,1)$ as $T\to\infty$, and $H_0^{\mathrm{MC}}$ is rejected at level $\alpha$ whenever $|Z_{G}|>z_{1-\alpha/2}$.
\end{proposition}

The variance estimator $\widehat\sigma^2$ correctly captures correlations within each draw (arising from the common partition), while independence across draws provides the large-sample theory. Since $|Z_G|$ grows mechanically with $T$ for any nonzero $D_G$, the permutation test of Section~\ref{subsec:permutation} is the primary substantive tool and the conditional $Z$-test serves as a Monte Carlo precision check.

\subsection{Permutation test: exactness and consistency}\label{subsec:permutation}

The conditional $Z$-test treats the kernel $G$ as the realised object and quantifies Monte Carlo precision. We complement it with a permutation test that fixes $G$ and randomises the demographic vector. The methodological precedent is \citet{chercye2023_approx_test}, who use a permutation test of approximate utility maximization in which the budgets are held fixed and consumption rays are permuted across observations, with the CCEI as test statistic. Our null and test statistic differ: $G$ encodes the population-level kernel structure produced by partition subsampling, and we shuffle the demographic vector $X$ to test whether observable characteristics organise that structure (rather than to test rationality itself). View the test statistic $D_G(X) = \overline G_{\mathrm w}(X) - \overline G_{\mathrm a}(X)$ as a function of the binary covariate $X = (X_1, \ldots, X_{|\mathcal I|})$ with $G$ held fixed. Sample $\sigma_1, \ldots, \sigma_B$ i.i.d.\ uniformly on the symmetric group $S_{|\mathcal I|}$ and define the two-sided permutation $p$-value
\[
\widehat p \;:=\; \frac{1 + \sum_{b=1}^{B} \mathbf 1\!\bigl\{|D_G(\sigma_b X)| \geq |D_G(X)|\bigr\}}{B + 1},
\qquad \sigma_b X := (X_{\sigma_b(1)}, \ldots, X_{\sigma_b(|\mathcal I|)}).
\]
The null hypothesis is exchangeability of $X$ given $G$:
\[
H_0^{\mathrm{perm}}:\ \Pr(X \in A \mid G) = \Pr(\sigma X \in A \mid G)\ \text{a.s. for every } \sigma \in S_{|\mathcal I|} \text{ and measurable } A.
\]

\begin{proposition}[Finite-sample validity]\label{prop:perm_exact}
Under $H_0^{\mathrm{perm}}$, for any $|\mathcal I|, T, B \geq 1$ and $\alpha \in (0,1)$, the permutation $p$-value is sub-uniform:
\[
\Pr_{H_0^{\mathrm{perm}}}\!\bigl(\widehat p \leq \alpha\bigr) \;\leq\; \frac{\lfloor \alpha (B+1) \rfloor}{B+1} \;\leq\; \alpha.
\]
\end{proposition}

\begin{proposition}[Consistency under oracle-kernel approximation]\label{prop:perm_consistent}
Suppose Assumption~\ref{ass:empirical_oracle_kernel} holds for some bounded oracle kernel $h_N^\Phi$. Let $\{(\theta_i, X_i)\}_{i=1}^{|\mathcal I|}$ be i.i.d.\ with $X_i \in \{0, 1\}$ and $\Pr(X_i = 1) = q \in (0, 1)$, and define
\[
\Delta^* \;:=\; \E\!\bigl[h_N^\Phi(\theta_i, \theta_j) \mid X_i = X_j\bigr] - \E\!\bigl[h_N^\Phi(\theta_i, \theta_j) \mid X_i \neq X_j\bigr].
\]
If $\Delta^* \neq 0$ and $B = B(|\mathcal I|, T) \to \infty$, then for any $\alpha \in (0, 1)$,
\[
\Pr\!\bigl(\widehat p \leq \alpha\bigr) \;\xrightarrow[\,|\mathcal I|, T, B \to \infty\,]{}\; 1.
\]
\end{proposition}

Proposition~\ref{prop:perm_exact} is a rank argument: under exchangeability of $X$ given $G$, the test statistics across the $B+1$ permutations are themselves exchangeable, so the observed value has uniform rank among them. Proposition~\ref{prop:perm_consistent} combines two ingredients: the observed statistic concentrates at $\Delta^* \neq 0$ (Assumption~\ref{ass:empirical_oracle_kernel} replaces $G_{i,j}$ by the oracle kernel at $o_p(1)$ cost, then a U-statistic law of large numbers and the continuous mapping theorem give convergence to $\Delta^*$), while the permutation null distribution concentrates at zero (mean exactly zero; variance $O(|\mathcal I|^{-1})$, Lemma~\ref{lem:perm_var}). Full proofs are in Appendix~\ref{app:perm_proofs}.

Two remarks. \emph{(i) The kernel is fixed.} Proposition~\ref{prop:perm_exact} requires nothing from $G$ beyond that the same realisation be used for $D_G(X)$ and the $D_G(\sigma_b X)$ values; the kernel can be biased, sparse, or under-dispersed without affecting the size guarantee. \emph{(ii) Exactness vs.\ consistency.} Proposition~\ref{prop:perm_exact} is unconditional: validity (sub-uniformity of $\widehat p$) holds regardless of how the kernel was generated. Consistency in Proposition~\ref{prop:perm_consistent} is the strictly stronger requirement that the realised kernel carry stable signal about latent types, which is what Assumption~\ref{ass:empirical_oracle_kernel} encodes. 


\subsection{Discrete covariates}

When $X_i\in\{1,\dots,L\}$, define for any two subsets $\mathcal G,\mathcal H\subset\{1,\dots,L\}$ (possibly equal)
\[
\mathcal P_{\mathcal G\!\parallel\!\mathcal H}=\{(i,j):i<j,\;(X_i\in\mathcal G\land X_j\in\mathcal H)\lor(X_i\in\mathcal H\land X_j\in\mathcal G)\}.
\]
For $\mathcal G = \mathcal H = \{g\}$, this is the within-category set $\{(i,j): i<j, X_i = X_j = g\}$; for $\mathcal G$ and $\mathcal H$ disjoint it is the symmetric union of cross-group pairs. Choosing $\mathcal G=\{g\}$ and $\mathcal H=\{g\}^{\mathrm c}$, the test assesses whether similarity within category $g$ exceeds similarity across categories:
\[
D_{G,g}^t = \frac{1}{m_{\{g\}\parallel\{g\}}}\sum_{(i,j)\in \mathcal{P}_{\{g\}\parallel\{g\}}}\delta_{i,j}^t - \frac{1}{m_{\{g\}\parallel\{g\}^{\mathrm c}}}\sum_{(i,j)\in \mathcal{P}_{\{g\}\parallel\{g\}^{\mathrm c}}}\delta_{i,j}^t,
\]
with empirical average $D_{G,g} = T^{-1}\sum_t D_{G,g}^t$ and across-draw variance $\widehat\sigma_g^2 = (T-1)^{-1}\sum_t (D_{G,g}^t - D_{G,g})^2$. Under the Monte-Carlo conditional null $H_0^{\mathrm{MC},g}: \E[D_{G,g}^t \mid \{D_i\}] = 0$ (implied by the population-level null that within-category and across-category mean co-typing coincide) and $\mathrm{Var}(D_{G,g}^t \mid \{D_i\}) > 0$, the statistic $Z_{G,g}=D_{G,g}\,/\,(\widehat{\sigma}_{g}/\sqrt{T})\xrightarrow{d}\mathcal N(0,1)$.

\subsection{Multiple covariates and conditional relevance}

Consider an $L$-dimensional vector of observables $X=(X^{1},\dots,X^{L})$. To isolate the partial contribution of $X^{k}$, we condition on $X^{-k}=(X^{1},\dots,X^{k-1},X^{k+1},\dots,X^{L})$. For every realisation $x^{-k}$, define the block $\mathcal I(x^{-k})=\{i\in\mathcal I:X_i^{-k}=x^{-k}\}$ and let $\mathcal B_k=\{x^{-k}:\mathcal I(x^{-k})\neq\varnothing\}$. Within each block $b\in\mathcal B_k$, define
\[
\mathcal P_{w,b}=\{(i,j):i<j,\,i,j\in\mathcal I(b),\,X^k_i=X^k_j\},\quad
\mathcal P_{a,b}=\{(i,j):i<j,\,i,j\in\mathcal I(b),\,X^k_i\neq X^k_j\},
\]
and, for blocks where both $m_{w,b}>0$ and $m_{a,b}>0$, the block-specific gap $D_{G,b}^t=m_{w,b}^{-1}\sum_{(i,j)\in\mathcal{P}_{w,b}}\delta_{i,j}^t - m_{a,b}^{-1}\sum_{(i,j)\in\mathcal{P}_{a,b}}\delta_{i,j}^t$. Let $\mathcal B_k^{\mathrm{valid}} := \{b \in \mathcal B_k : m_{w,b} > 0,\, m_{a,b} > 0\}$, and assume $\mathcal B_k^{\mathrm{valid}} \neq \varnothing$ (this requires at least one conditioning stratum that contains both $X^k=0$ and $X^k=1$ agents \emph{and} at least two agents in each label class; the weaker condition that $X^k$ is not a deterministic function of $X^{-k}$ guarantees label variation within some stratum but not enough pairs of each type). Aggregating across valid blocks with weights renormalised to sum to one:
\[
D_{G}^t(k)=\frac{\sum_{b\in\mathcal B_k^{\mathrm{valid}}}|\mathcal I(b)|\,D_{G,b}^t}{\sum_{b\in\mathcal B_k^{\mathrm{valid}}}|\mathcal I(b)|},
\qquad
D_{G}(k)=\frac{1}{T}\sum_{t=1}^{T}D_{G}^t(k),
\qquad
\widehat\sigma(k)^2 = \frac{1}{T-1}\sum_{t=1}^T \bigl(D_G^t(k) - D_G(k)\bigr)^2.
\]
The conditional statistic $Z_{G}(k)=D_{G}(k)\,/\,(\widehat{\sigma}(k)/\sqrt{T})\xrightarrow{d}\mathcal N(0,1)$ under the Monte-Carlo conditional null $\E[D_G^t(k) \mid \{D_i\}] = 0$ (implied by the population-level $H_0:\overline\pi_{w\mid -k}=\overline\pi_{a\mid -k}$, where the population contrasts are averaged across valid blocks weighted by block size), provided $\mathrm{Var}(D_G^t(k) \mid \{D_i\}) > 0$.

\paragraph{Conditional permutation validity.}
The analogous conditional permutation test shuffles $X^k$ within each stratum defined by $X^{-k}$ and reports the resulting permutation $p$-value. Validity follows from applying Proposition~\ref{prop:perm_exact} stratum-by-stratum: under the conditional null that $X^k \mid X^{-k}$ is exchangeable within each stratum, the per-stratum statistic $D_{G,b}(\sigma X^k)$ has the same conditional distribution as the observed $D_{G,b}(X^k)$ for any within-stratum permutation $\sigma$. The aggregated statistic $D_G(k)$ is a deterministic function of the per-stratum statistics, so the rank argument of Proposition~\ref{prop:perm_exact} delivers the sub-uniform bound $\Pr(\widehat p \leq \alpha) \leq \alpha$ in finite samples, for any $|\mathcal I|, T, B \geq 1$.

\paragraph{Discussion.}
The mean-difference framework tests whether an observable covariate is associated with the metric structure of the population. In the conditional version, $D_G(k)$ is the average within-stratum kernel gap between agents who share $X^k$ and those who do not, after conditioning on the other observables. A positive and significant $Z_G(k)$ indicates a conditional association between $X^k$ and the kernel structure that is not explained by the remaining demographics; under the design, this is a descriptive partial-association statement, not a causal-attribution claim.

The tests as developed require discrete covariates to define within- and across-group pair sets. Continuous covariates (e.g., income measured in dollars) do not provide natural groupings; applying our tests requires discretising them into categories, which introduces arbitrary thresholds and some loss of information. In the application, all demographics are recorded in ordered categories, so this limitation does not bind. An alternative for continuous covariates would be to regress the kernel entries $G_{i,j}$ directly on pairwise functions of demographic distances, bypassing the group-based framework.

The conditional test evaluates each covariate's marginal contribution holding others fixed, but does not test interaction effects (e.g., whether the joint effect of age and family size exceeds the sum of their individual effects). Incorporating interactions is straightforward in principle - one can define pair sets based on joint category membership - but requires larger samples to avoid sparse cells.

\subsection{Spectral attribution of the test statistic}\label{subsec:spectral_attribution}

The eigendecomposition of $\Pi$ introduced in Section~\ref{subsec:kernel_discussion} attributes the population mean-difference contrast $\overline\pi_{\mathrm w} - \overline\pi_{\mathrm a}$ to specific dimensions of preference heterogeneity. Since $\pi_{i,j}=\sum_\ell \lambda_\ell\, v_\ell(i)\,v_\ell(j)$, the contrast decomposes as
\begin{equation}\label{eq:DG_decomp}
\overline\pi_{\mathrm w} - \overline\pi_{\mathrm a} \;=\; \sum_{\ell=1}^{|\mathcal I|} \lambda_\ell\, D_\ell,
\qquad
D_\ell \;=\; \underset{X_i=X_j}{\text{mean}}\; v_\ell(i)\,v_\ell(j)
        \;-\; \underset{X_i\neq X_j}{\text{mean}}\; v_\ell(i)\,v_\ell(j),
\end{equation}
where $D_\ell$ measures the alignment of eigenvector $v_\ell$ with the demographic~$X$. The product $\lambda_\ell D_\ell$ is the contribution of the $\ell$-th preference axis to the population contrast, and $\lambda_\ell D_\ell / (\overline\pi_{\mathrm w} - \overline\pi_{\mathrm a})$ is its share. This decomposition is exact at the population level. The empirical analogue plugs in the eigendecomposition of $G$ in place of $\Pi$ and substitutes $D_G$ for $\overline\pi_{\mathrm w} - \overline\pi_{\mathrm a}$; the decomposition then reveals \emph{which dimensions of preference heterogeneity a given demographic captures} in the realised kernel: a demographic that aligns with the leading eigenvector captures the dominant source of heterogeneity, while one whose signal is dispersed across many small eigenvalues captures only peripheral variation.

\section{Simulation}\label{sec:simulation}

This section evaluates the method in a controlled environment with known preferences. We compare the pairwise benchmark $\rho_{i,j}$ with the partition kernel and exhibit the joint-rationality gap; we show that the kernel responds to type structure rather than rationality per se; we document the by-type kernel gradient and verify the contrast-rank condition required by Theorem~\ref{thm:finite_type_witness}; we report the mean-difference test across designs; and we point to additional diagnostics in the appendix. 

\subsection{Data-generating process}

We consider $I=100$ consumers, each observed at $N_i=200$ independent budgets. Consumer $i$ maximises
\[
u_i(q_1,q_2)=q_1^{\alpha_i}\,q_2^{1-\alpha_i},\qquad\text{s.t. } p_{i,1}^n q_1+p_{i,2}^n q_2\leq 1.
\]
For the mean-difference tests, there are two latent types $\alpha_i\in\{\alpha^{\mathrm{low}},\alpha^{\mathrm{high}}\}$, with $(\alpha^{\mathrm{low}},\alpha^{\mathrm{high}})\in\{(0.2,0.8),(0.4,0.6)\}$, and a binary observable $X_i$ correlated with the type through $\Pr(\alpha_i=\alpha^{\mathrm{high}}\mid X_i=1)=\gamma$, with $\gamma\in\{0.5,0.6,0.8\}$. The case $\gamma=0.5$ serves as a null: the covariate is independent of the latent type. For the structural calibration and power analysis, we use five types $\alpha_i\in\{0.20,0.35,0.50,0.65,0.80\}$ with equal allocation ($20$ agents per type). Prices are drawn from $[p_{\min},p_{\max}]^2$, contrasting $[0.5,5]$ (wide) with $[0.5,1.5]$ (tight).

Unless otherwise noted, we use $T=100$ Monte Carlo draws and the scalable sequential partition, implemented in \texttt{R} and running in parallel on a server with 64~CPU cores. No optimisation solver is required.

\subsection{Pairwise benchmark and joint-rationality gap}\label{subsec:sim_calibration}

\begin{figure}[ht]
\centering
\includegraphics[width=0.78\textwidth]{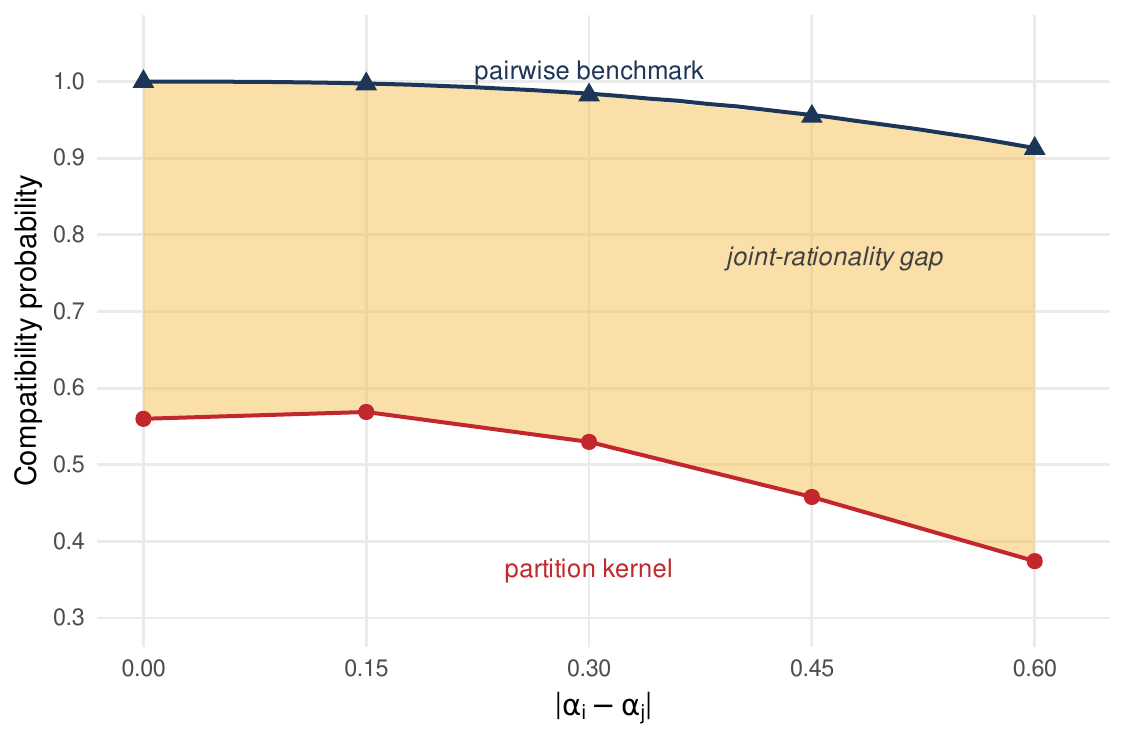}
\caption{Pairwise benchmark, partition kernel, and joint-rationality gap as a function of the taste gap $|\alpha_i-\alpha_j|$, Cobb--Douglas preferences with wide prices $[0.5,5]^2$. The top (blue) curve plots the pairwise compatibility $\rho_{i,j}$ computed from $500{,}000$ independent price draws per point; the bottom (red) curve plots the partition-kernel mean $\bar\pi_{i,j}$ from the five-type simulation at $N=1$ ($I=100$, $T=100$, sequential partition), averaged across all agent pairs with the same taste gap. The shaded area is the joint-rationality gap $\Delta_{i,j}=\rho_{i,j}-\pi_{i,j}$.}
\label{fig:rho_vs_distance}
\end{figure}

Figure~\ref{fig:rho_vs_distance} makes the two ingredients of the logical architecture of Section~\ref{subsec:pairwise} concrete. The upper (blue) curve is the pairwise benchmark $\rho_{i,j}$ -- the probability that two Cobb--Douglas agents are pairwise GARP-consistent -- computed from $500{,}000$ independent price draws per point. The lower (red) curve is the partition-kernel mean $\bar\pi_{i,j}$ from the five-type simulation at $N=1$, averaged across all agent pairs with the same taste gap $|\alpha_i-\alpha_j|$. The shaded area between them is the joint-rationality gap $\Delta_{i,j}=\rho_{i,j}-\pi_{i,j}$, the quantity that the partition adds over pairwise compatibility.

The pairwise benchmark behaves as predicted: Proposition~\ref{prop:CD_no_false} guarantees $\rho_{i,j}=1$ when $\alpha_i=\alpha_j$, and the simulation traces a $\rho_{i,j}$ that is numerically monotonically decreasing in $|\alpha_i-\alpha_j|$. Yet $\rho_{i,j}$ remains close to~$1$ throughout -- even at the widest taste gap $|\alpha_i-\alpha_j|=0.60$, pairwise GARP is violated in fewer than $9\%$ of draws, and the total range of $\rho_{i,j}$ across the grid is under $0.09$. The pairwise check with $N=1$, which reduces to the Weak Axiom for two observations, is therefore close to uninformative in two goods.

The joint-rationality gap supplies the missing discriminatory power. Its magnitude is striking: $\Delta_{i,j}$ ranges from $0.44$ at zero taste gap to $0.54$ at the widest gap -- $\Delta$ values are five to six times the total range of $\rho_{i,j}$ across the same grid. The partition is not a marginal refinement of the pairwise check; it is doing essentially all of the separation. Moreover, $\Delta_{i,j}$ widens with the taste gap, so higher-order revealed-preference constraints across the population cut more sharply precisely where preferences are most different.

\subsection{Structure, not rationality per se}\label{subsec:sim_power}

\begin{figure}[ht]
\centering
\includegraphics[width=0.75\textwidth]{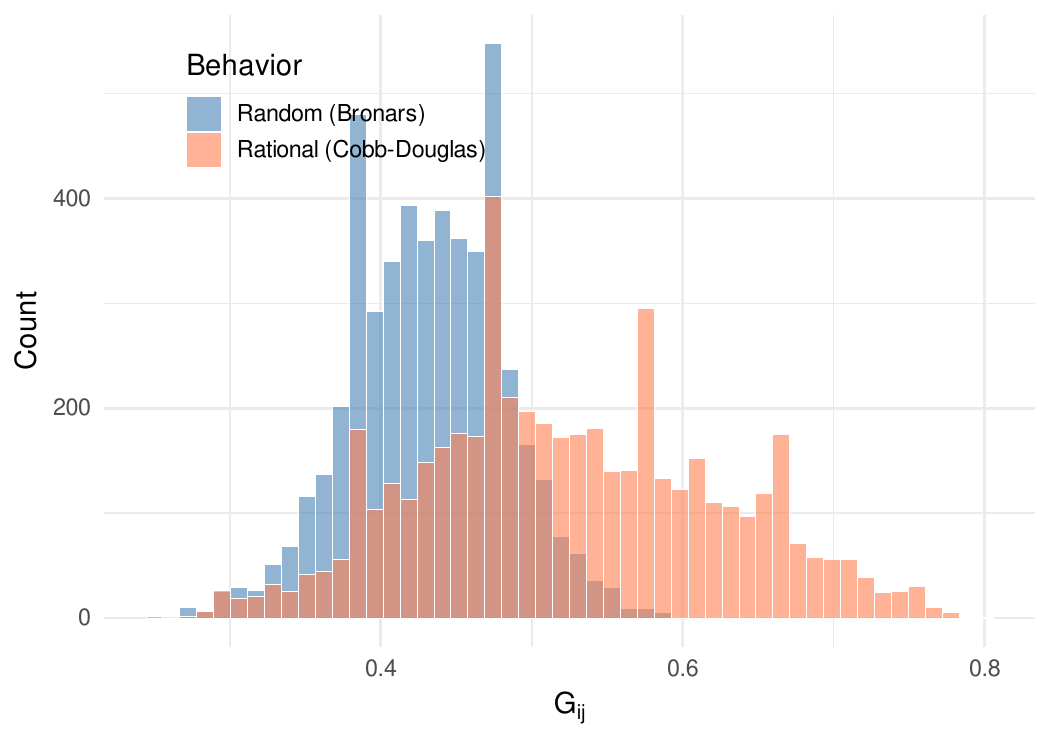}
\caption{Distribution of $G_{i,j}$ under rational behavior (5 Cobb--Douglas types) and random behavior (Bronars benchmark). Same budgets in both conditions; $I=100$, $T=100$, sequential partition. Off-figure summary statistics: rational distribution has mean $0.52$ and range $[0.25, 0.80]$; random distribution has mean $0.43$ and range $[0.25, 0.62]$. The rational right tail above $0.62$ corresponds to same-type pairs.}
\label{fig:bronars}
\end{figure}

To verify that the kernel detects structure rather than rationality per se, we compare $G_{i,j}$ under two conditions using the same population of budgets: rational Cobb--Douglas optimisation and a Bronars benchmark \citep{bronars} in which each agent draws a bundle uniformly at random from their budget simplex. Figure~\ref{fig:bronars} displays the distributions. Under rational heterogeneous behavior, $G_{i,j}$ is dispersed (mean $0.52$, range $0.25$--$0.80$), with a distinctive right tail of same-type pairs above $0.62$. Under random behavior, $G_{i,j}$ concentrates at moderate levels (mean $0.43$, range $0.25$--$0.62$) with no high-similarity tail. The right tail of the rational distribution has no counterpart under randomness, confirming that the method detects structured preference heterogeneity rather than individual rationality.

\subsection{Contrast-rank condition}\label{subsec:sim_kernel}\label{subsec:sim_witness}


Theorem~\ref{thm:finite_type_witness} relies on the contrast-rank condition~\eqref{eq:CRC}: the type-level kernel $S_N^\Phi$ has full rank on the $(K-1)$-dimensional contrast subspace. Because the simulation provides ground-truth type labels, we can check~\eqref{eq:CRC} directly from Table~\ref{tab:type_means}. On the five-type Cobb--Douglas grid at $N=1$, $T=100$, $I=100$, the empirical type-level kernel has all four contrast-space eigenvalues strictly positive ($0.098$, $0.075$, $0.0036$, $0.0014$). The two leading eigenvalues capture most of the type structure (low- vs.\ high-$\alpha$, central- vs.\ extreme-$\alpha$); the smaller two reflect finer adjacent-type distinctions. The contrast-rank condition therefore holds in the canonical worked case, so spectral clustering on the realised kernel recovers the latent five-type partition with vanishing misclassification as $|\mathcal I|, T \to \infty$ (Theorem~\ref{thm:finite_type_witness}).

\subsection{Mean-difference test results}\label{subsec:sim_test}

Table~\ref{tab:DG_four_panels} reports $D_G$ for $T=100$ across twelve two-type designs (two taste gaps, three alignment levels including $\gamma=0.5$ as a null, two price ranges) using the sequential partition at $N\in\{1,3,5,10,20\}$. The cost--power tradeoff is visible across $N$. At $N=1$, only the strongest configurations are significant: in Panel~A, the large-taste-gap strongly-aligned design ($D_G = 0.028$); in Panel~B, the same design with wider effect ($D_G = 0.066$) plus the large-taste-gap moderately-aligned design ($D_G = 0.006$). The moderate taste gap ($\alpha^{\mathrm{low}}=0.4$, $\alpha^{\mathrm{high}}=0.6$) is entirely undetectable at $N=1$ under either price regime, but the same design under strong alignment ($\gamma=0.8$) rises from $D_G=0.001$ at $N=1$ to $0.020$ at $N=3$, $0.076$ at $N=5$, $0.262$ at $N=10$, and $0.392$ at $N=20$ (Panel~A); the big jump happens between $N=5$ and $N=10$. The interaction between $N$ and price variation is informative: with the large taste gap under moderate alignment ($\gamma=0.6$), narrow prices push $D_G$ to $\sim 0.030$ at $N\geq 10$, about three times the wide-price counterpart ($\sim 0.010$); narrow-price discrimination is sharper because concentrated price variation makes GARP constraints more frequently binding. Under the null ($\gamma=0.5$), $D_G$ at $N=1$ is indistinguishable from zero at conventional levels in both panels (with one marginal 10\%-significance exception in Panel~B for the wide-taste-gap null, consistent with sampling noise at the smallest scale); at larger $N$ the $Z$-statistic grows mechanically with cross-draw precision and produces nominal significance even for tiny null-deviations, consistent with the framing in Section~\ref{sec:tests} that this asymptotic $Z$-test is a Monte Carlo precision check while the permutation analysis carries the substantive inference. Larger $N$ progressively uncovers finer preference heterogeneity at greater computational cost.


\subsection{Additional diagnostics and the role of the simulation design}\label{subsec:sim_diagnostics}\label{subsec:sim_design_choice}

Four further properties of the kernel are documented in Appendix~\ref{app:simulation_details}.
\begin{enumerate}
\item \emph{Permutation-test size and power} (Table~\ref{tab:perm_power}). At $I=100$, $T=100$, $B=1{,}000$, $R=200$ outer replications: empirical size under the null ($\gamma=0.5$) is $0.055$ at $\alpha=0.05$, within the binomial band around nominal; power rises to $35.5\%$ at the moderate alternative ($\gamma=0.6$, $\bar D_G \approx 0.4$ pp) and to $1.000$ at the strong alternative ($\gamma=0.8$, $\bar D_G \approx 4$ pp).
\item \emph{Discrimination ratio across $N$} (Table~\ref{tab:N_spectrum}). The ratio of mean same-type to mean different-type similarity rises monotonically from $1.07$ at $N=1$ to over $500$ at $N=25$, as each additional within-agent observation imposes one more cross-pair rationality constraint when distinct types are pooled.
\item \emph{Continuous-heterogeneity smoothing} (Figure~\ref{fig:continuous_heterogeneity}). With $\alpha_i \sim \mathrm{Uniform}(0.1, 0.9)$, the mean kernel declines smoothly from $0.55$ at zero gap to $0.28$ at the largest gap at $N=1$; by $N=10$, $67\%$ of pairs have $G<0.1$. The bandwidth role of $N$ is explicit: small $N$ behaves like a smoothing kernel over the latent parameter space, large $N$ like a sharp type-indicator.
\item \emph{Rule-dependence diagnostics} (Table~\ref{tab:milp_comparison}). The four partition rules produce kernels that are moderately correlated at the pair level (Pearson $r = 0.73$ to $0.80$ across pairs of rules), highly correlated globally (Frobenius cosine $\approx 0.99$), and well aligned on their leading eigenvectors ($|\mathrm{cor}(v_1)| = 0.82$ to $0.86$). Aggregate-level claims the paper makes are credible across rules; specific contrast-level significances for smaller demographic effects can still shift across rules (Table~\ref{tab:robustness_demo}).
\end{enumerate}


\section{Hidden Preference Heterogeneity in Scanner Data}\label{sec:application}

Pairwise GARP classifies almost every household pair in the scanner panel as mutually compatible; the partition kernel reveals that the population is in fact much more heterogeneous than this near-saturation suggests. We apply the methodology to the \emph{Stanford Basket} scanner panel used by \citet{bell1998}, \citet{shum2004}, \citet{hendel2006,hendel2006_b}, and \citet{echenique2011}. The data track grocery purchases of households shopping in four Midwestern U.S.\ supermarkets over $26$ consecutive months (June 1991--June 1993), classified into $379$ food categories. Prices are common across households within each month, so the kernel reflects preference heterogeneity rather than heterogeneity in the price environment. 

We use all $480$ households with demographic information from the replication package of \citet{echenique2011}. Four demographic variables are available (Table~\ref{tab:summary_stat_scanner_data}): family size, income, age, and education, each in three ordered categories. We encode them as twelve binary dummies $X^{k}\in\{0,1\}$ ($k=1,\dots,12$) for the dummy-level mean-difference contrasts. For the conditional tests, the conditioning set is the three remaining \emph{demographic variables} (each with three categories, yielding $3^3 = 27$ strata), not the eleven other dummies --- conditioning a dummy on the other two dummies of the same demographic would be degenerate by construction.

For each household $i\in\mathcal I$ and month $n$, let $q_i^n\in\mathbb R_+^{379}$ be the quantity vector and $p_i^n$ the price vector. We run $T=500$ Monte Carlo draws using the scalable sequential partition, which handles the full sample of $480$ households in under twenty minutes on a $64$-core server. We recover the similarity matrix $G=\{G_{i,j}\}$ --- the unbiased Monte Carlo estimator of the population kernel $\Pi=\{\pi_{i,j}\}$ --- and compute, for each demographic dummy, the mean-difference $D_G$ and its standardized test statistic $Z_G = D_G / (\widehat\sigma/\sqrt T)$.

\subsection*{What the kernel reveals and what it measures}

\paragraph{The joint-rationality gap.}
Section~\ref{subsec:sim_calibration} documented the joint-rationality gap $\Delta_{i,j}=\rho_{i,j}-\pi_{i,j}$ under Cobb--Douglas preferences in simulation. We compute the same object directly on the scanner data at $N=1$ (the main full-sample baseline) and $N=5$. For each of the $\binom{480}{2}\approx 115{,}000$ household pairs, $\rho_{i,j}$ is the probability that a randomly drawn pool of $2N$ observations satisfies pairwise GARP, and $\pi_{i,j}$ is the partition-kernel entry $G_{i,j}$.

\begin{figure}[ht]
\centering
\includegraphics[width=0.95\textwidth]{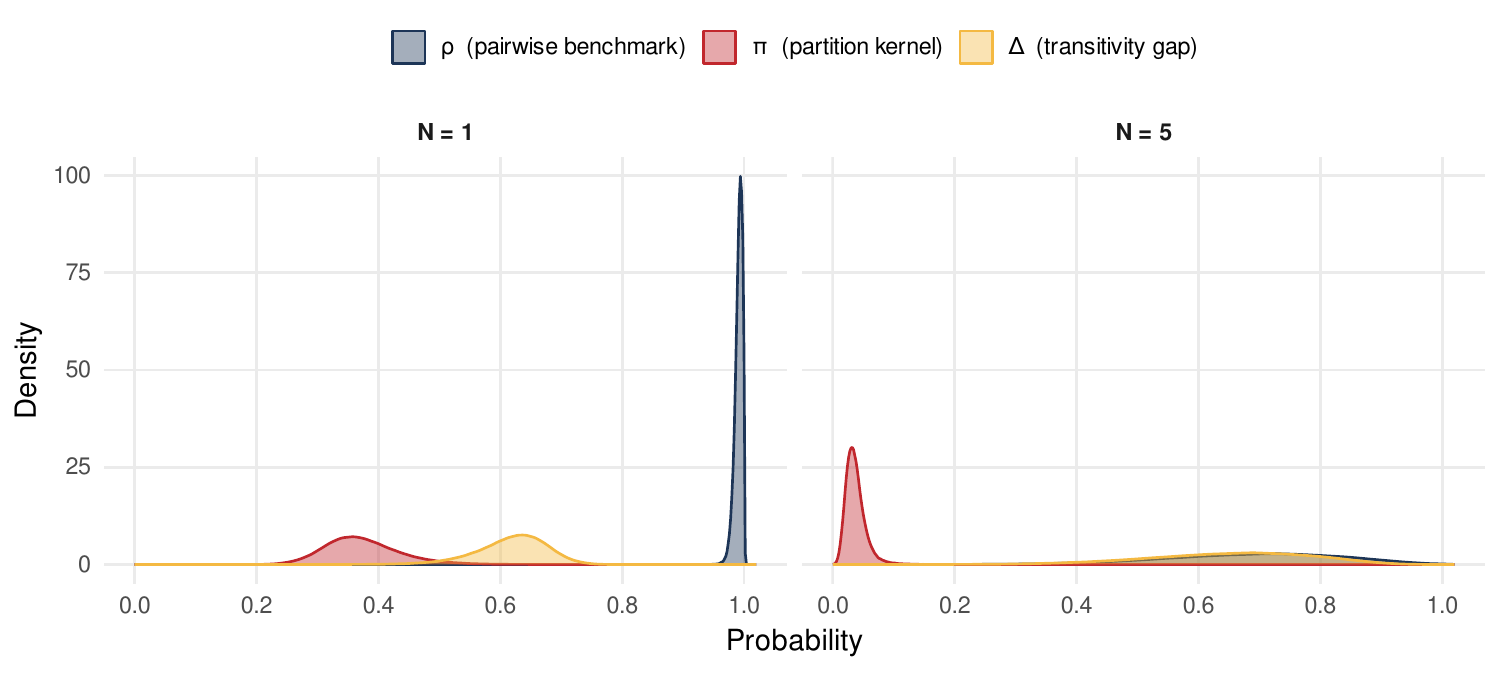}
\caption{Joint-rationality gap on the scanner data at $N=1$ (left) and $N=5$ (right). Each panel plots the densities across all $\binom{480}{2}=114{,}960$ household pairs of the pairwise benchmark $\rho_{i,j}$, the partition-kernel entry $\pi_{i,j}$ from a full-sample run at the relevant $N$ ($T=500$ draws), and the joint-rationality gap $\Delta_{i,j}=\rho_{i,j}-\pi_{i,j}$. At $N=1$, $\rho_{i,j}$ is computed exactly from all $N_i\times N_j$ observation combinations; at $N=5$, it is estimated by Monte Carlo (100 draws per pair, 5 observations from each agent). Summary statistics computed off-figure: at $N=1$, $\bar\rho=0.991$ (95\% of pairs have $\rho_{i,j}\geq 0.98$; 7\% have $\rho_{i,j}=1$), $\bar\pi=0.373$, $\bar\Delta=0.618$ (97\% of pairs have $\Delta_{i,j}>0.5$); at $N=5$, $\bar\rho=0.689$ (0.3\% have $\rho_{i,j}=1$ across all draws), $\bar\pi=0.037$, $\bar\Delta=0.652$ (87\% have $\Delta_{i,j}>0.5$). In the right panel, both $\rho$ and $\pi$ decay with $N$, but $\pi$ decays faster, leaving $\Delta$ large.}
\label{fig:scanner_transitivity}
\end{figure}

At $N=1$, the pairwise benchmark concentrates tightly near one (mean $0.991$; $7\%$ of pairs have $\rho_{i,j}=1$, meaning no observation pair between them violates pairwise GARP). A test restricted to pairwise compatibility would therefore find the scanner households almost indistinguishable. The partition-kernel estimator $G_{i,j}$ of $\pi_{i,j}$ is much lower (mean $0.373$, roughly $38\%$ of the pairwise benchmark): the population-level GARP constraints embedded in the partition separate households even when bilateral compatibility cannot. The joint-rationality gap is accordingly large (mean $\bar\Delta=0.618$; $97\%$ of pairs have $\Delta_{i,j}>0.5$, where $\Delta_{i,j}$ here is the plug-in estimate $\rho_{i,j} - G_{i,j}$). The gap aggregates two components --- preference-driven separation between households and mechanical contributions (rule-specific path-dependence, third-agent conflicts, and finite budget overlap); the Cobb--Douglas simulation in Section~\ref{subsec:sim_calibration} quantifies the mechanical baseline (which alone produces $\bar\Delta \approx 0.44$ at zero taste gap), so the empirical $0.62$ is above this baseline rather than a pure measure of preference separation. At $N=5$, the partition kernel collapses to a mean of $0.037$ while the pairwise benchmark falls only to $0.689$, so the gap remains substantial ($\bar\Delta=0.652$). Figure~\ref{fig:scanner_transitivity} plots the densities at both $N$. The partition-based kernel adds the bulk of the discriminatory signal: the demographic tests, the spectral structure, and the robustness checks all operate on the population-level transitivity that the kernel encodes, not on the near-saturated pairwise benchmark.

\paragraph{Spectral decomposition.}
On the doubly centered kernel (which removes the level effect dominating the raw matrix), the leading eigenvalue ratio is $\lambda_1/\lambda_2 = 3.2$, suggesting a small number of genuine heterogeneity dimensions. The first principal coordinate $v_1$ has correlation $-1.000$ with the row mean of $G$ and therefore captures, up to sign, the average compatibility of household $i$ with the rest of the population. Reading this spectral structure as evidence of latent finite types via Theorem~\ref{thm:finite_type_witness} requires Assumption~\ref{ass:empirical_oracle_kernel}: the partition algorithm by itself does not guarantee that the realised kernel approximates an oracle type-level structure, only that aggregate co-typing probabilities are well-defined; spectral recovery is a statement about $G$ relative to such an oracle, not about $G$ in isolation. 

\subsection*{Demographic correlates}

\paragraph{Inference and the Afriat-efficiency path.}
Table~\ref{tab:cond_demo} reports the conditional mean-difference statistic for each demographic dummy at $N \in \{1, 3, 5\}$ and at the $e$-GARP relaxation $e = 0.95$; substantive inference is anchored on the permutation tests below since the conditional $Z$-statistics are mechanically large at $T=500$ draws. The largest single effect is large family size ($D_G = -3.8$ pp at $N=5$); young and old age, and high and low income, contribute at the $-2$ pp level; education contrasts and mid-category dummies are an order of magnitude smaller. The headline contrasts retain sign and rank across $N$ and across the $e$-GARP relaxation. Table~\ref{tab:egarp_path} extends the analysis to a full Afriat-efficiency path $e \in \{1, 0.99, 0.97, 0.95, 0.90\}$: pairwise checks become uninformative as $e$ falls ($\bar\rho$ rises from $0.69$ to $0.97$), while the partition kernel preserves a substantial joint-rationality gap ($\bar\Delta=0.68$ at $e=0.90$), and the headline family-size and income effects strengthen modestly under relaxation. 


\paragraph{Permutation validation.}
We complement the $Z$-tests with two permutation analyses, both computed on the partition kernel at $N=1$ where pair-wise variance is largest: an unconditional permutation (Table~\ref{tab:permutation} in Appendix~\ref{app:additional_tables}) and a conditional permutation (Table~\ref{tab:cond_permutation}).\footnote{See Appendix~\ref{app:simulation_details} for the choice of $N=1$ for permutation and the test's finite-sample power calibration (Table~\ref{tab:perm_power}: $35.5\%$ rejection at $\alpha=0.05$ against a moderate alternative).} Unconditional permutation confirms that family size, age, and income are genuine at the unconditional level ($p<0.03$); education is not. Conditional permutation, which shuffles each variable within strata of the others, leaves only high income marginally significant ($p=0.027$); family size, age, and the rest become statistically insignificant. Applying the \citet{benjamini1995} false-discovery-rate (FDR) adjustment across the twelve dummies leaves no conditional contrast significant at $q<0.10$, while the headline unconditional contrasts survive FDR correction at conventional levels. The substantive headline is therefore disciplined: demographics have detectable unconditional associations with the kernel structure but near-null partial explanatory power. 

\paragraph{Robustness across partition rules.}
Table~\ref{tab:robustness_demo} compares the scalable sequential partition used as the baseline (which places each agent into the first existing block whose pooled data remains GARP-consistent) with three variants of a MILP-based partition that solves an optimisation problem at each step under different tie-breaking conventions (min-priority, max-priority, random). The two leading contrasts in absolute size are rule-stable in sign and significance: Large family size and Young age are negative and significant under all four rules. High income is significant under sequential and MILP-min and attenuates (same sign) under MILP-max and MILP-random. Three patterns are rule-sensitive. First, two effects (Mid family size, Old age) are significant only under the sequential rule and attenuate to non-significance under all three MILP variants; for Old age the sequential-rule magnitude is comparable to Large family size, so its rule-sensitivity is substantive rather than marginal. Second, two education contrasts (Primary, College) flip sign across rules: Primary is positive and significant under sequential and MILP-min but turns slightly negative under MILP-max; College is negative under sequential but positive under all three MILP variants (significant under MILP-max and MILP-random). Third, High School is negative and significant under sequential and MILP-random but loses significance under MILP-min and MILP-max. The aggregate-level summary --- the leading centred eigenvector and the ranking of the two largest contrasts --- is rule-stable; specific contrast-level significances are rule-indexed (\S\ref{subsec:rule_dependence} and Appendix~\ref{app:simulation_details}). 

\section{A Risky-Choice Application}\label{sec:lottery}

The construction extends to any choice environment with a testable consistency condition (Section~\ref{subsec:kernel_discussion}). We use this portability to do something the scanner application is not designed to showcase: trace the kernel's $N$-progression empirically. We use the Choice Prediction Competition 2018 (CPC18) dataset \citep{cpc18_dataset}, which builds on the experimental framework of \citet{erev2017}, restricted to the first no-feedback trial with binary lotteries and no ambiguity. After cleaning, $686$ subjects make a total of $10{,}243$ binary choices (mean $14.9$ per subject, range $5$--$24$) over $161$ unique lotteries. Each choice is a directed edge in a population-level lottery graph. A pool of choices from one or more subjects is jointly rationalisable by a monotone utility consistent with first-order stochastic dominance (FSD) iff the union of those edges and the FSD-implied edges is acyclic. The partition rule places each subject into the first existing block whose pooled graph remains acyclic when augmented. The pairwise compatibility $\rho_{i,j}$ is computed exactly by enumeration over the $n_i \times n_j$ choice combinations per pair.

Table~\ref{tab:cpc18_summary} reports the headline numbers. The pairwise benchmark $\bar\rho = 0.959$, computed at $N=1$ (one binary choice from each subject) and held as a fixed bilateral baseline throughout, is again near-saturated, as in the scanner application; the partition kernel does the discriminatory work, with a gap of $0.38$ relative to this baseline already at $N=1$. As $N$ rises from $1$ to $5$ to $10$, the mean partition kernel collapses from $0.582$ to $0.092$ to $0.024$ (a factor of $24$). The reported gaps $\bar\Delta_N = \bar\rho_{N=1} - \bar\pi^\Phi_N = 0.38, 0.87, 0.94$ at $N=1, 5, 10$ are therefore comparisons to the $N=1$ pairwise baseline rather than to the same-$N$ pairwise benchmark, which would require recomputing $\rho$ at each $N$ over the $\binom{n_i}{N}$ subsets per subject. The centred eigenvalue ratio $\lambda_1/\lambda_2$ drops from $4.14$ to $2.12$ to $1.18$. The $N=1$ ratio of $4.14$ is comparable to the scanner value of $3.2$, where a single typicality axis dominates; by $N=10$ no single component does. Figure~\ref{fig:cpc18_transitivity} visualises the same progression as densities. As in the scanner case, the reported gap reflects both preference-driven separation between subjects and the mechanical components documented in Section~\ref{subsec:sim_calibration}.


\section{Conclusion}\label{sec:conclusion}

This paper develops a continuous nonparametric metric of consumer heterogeneity from joint rationalisability of revealed-preference data. Pairwise rationality tests are well-suited to assessing individual rationality but do not directly deliver a measure of how much consumers differ at the population level, and applied questions about heterogeneity have therefore often relied on parametric demand systems where distance is read off the parameter space. The metric we propose is continuous, population-endogenous, and built from the rationality structure of the data itself, with a Hilbert-space embedding and a spectral structure that recovers latent preference types under a checkable contrast-rank condition.

The construction proceeds by repeatedly subsampling observations and partitioning agents into rationality-consistent blocks. The resulting rule-indexed similarity matrix is positive semi-definite, defines a Hilbert-space pseudometric with a triangle inequality, and decomposes the population's heterogeneity into orthogonal dimensions. Under an oracle-kernel approximation and a checkable type-separation (contrast-rank) condition on the type-level partition kernel, spectral clustering on the realised matrix recovers the latent type partition with vanishing error rate as the population and Monte Carlo draw count grow; the result extends to approximate optimisation. Inference on whether demographics organise the kernel structure proceeds via two complementary tests: a Monte-Carlo-conditional test and a permutation test that we show is exact in finite samples and consistent whenever the demographic-type association is detectable through the kernel mean-difference contrast.

A controlled Cobb--Douglas simulation with known preference types isolates the partition mechanism: it verifies the contrast-rank condition numerically and shows that the kernel separates structured rational behaviour from a Bronars-style random benchmark. The simulation establishes that the mechanism the theory predicts is the mechanism the kernel exhibits.

The scanner application uses grocery data on $480$ US households: pairwise compatibility is near-saturated and the partition kernel does essentially all the discriminatory work, with standard demographics explaining only a modest part of the kernel structure after stratification (consistent with \citet{crawford2012}). The lottery application uses $686$ subjects making binary lottery choices and traces the kernel's response to within-subject information directly: as more choices per subject become available, mean co-typing collapses, the joint-rationality gap saturates, and the kernel's spectral structure resolves from a single dominant axis into multiple comparable axes. We do not interpret the lottery results as recovering risk-preference types; the validation shows that the construction's mechanism carries over to a different choice environment with a different rationality criterion.

The mathematical construction - partition kernel, PSD property, and Hilbert-space distance - extends to any environment with a testable consistency condition: intertemporal choice, expected-utility tests, cost-minimisation for firms. We emphasise that this portability is established for the kernel construction and its distance structure; identification, power, and the interpretation of the resulting spectrum must be re-established for each new criterion, as we do separately here for the GARP-based scanner application and the FSD-based lottery application. A scalable partition procedure that requires no optimisation software makes the framework practical for large populations. Natural next steps include direct out-of-sample validation as a segmentation device for demand, and sharper finite-sample identification results.

\bibliographystyle{plainnat}
\bibliography{bibliography}

\clearpage

%

\clearpage
\newpage
\section*{Tables}


\begin{table}[ht]
\caption{Mean--difference statistic $D_G$, $T=100$, sequential partition.}
\label{tab:DG_four_panels}
\centering
\small
\begin{tabular}{llccccc}
\toprule
Alignment & Taste gap & $N=1$ & $N=3$ & $N=5$ & $N=10$ & $N=20$ \\
\midrule
\multicolumn{7}{l}{\emph{Panel A: Price support $[0.5,\,5]^2$}} \\
$\gamma=0.8$ & $(0.2,\,0.8)$ & 0.0281$^{***}$ & 0.1840$^{***}$ & 0.2344$^{***}$ & 0.2424$^{***}$ & 0.2424$^{***}$ \\
$\gamma=0.8$ & $(0.4,\,0.6)$ & 0.0006 & 0.0196$^{***}$ & 0.0757$^{***}$ & 0.2623$^{***}$ & 0.3917$^{***}$ \\
$\gamma=0.6$ & $(0.2,\,0.8)$ & 0.0011 & 0.0064$^{***}$ & 0.0085$^{***}$ & 0.0097$^{***}$ & 0.0097$^{***}$ \\
$\gamma=0.6$ & $(0.4,\,0.6)$ & 0.0009 & 0.0020 & 0.0074$^{***}$ & 0.0113$^{***}$ & 0.0150$^{***}$ \\
$\gamma=0.5$ (null) & $(0.2,\,0.8)$ & -0.0004 & -0.0061$^{***}$ & -0.0079$^{***}$ & -0.0082$^{***}$ & -0.0082$^{***}$ \\
$\gamma=0.5$ (null) & $(0.4,\,0.6)$ & -0.0006 & -0.0000 & -0.0011 & -0.0022$^{***}$ & -0.0038$^{***}$ \\
\addlinespace
\multicolumn{7}{l}{\emph{Panel B: Price support $[0.5,\,1.5]^2$}} \\
$\gamma=0.8$ & $(0.2,\,0.8)$ & 0.0662$^{***}$ & 0.2683$^{***}$ & 0.3198$^{***}$ & 0.3296$^{***}$ & 0.3296$^{***}$ \\
$\gamma=0.8$ & $(0.4,\,0.6)$ & 0.0013 & 0.0482$^{***}$ & 0.1452$^{***}$ & 0.2477$^{***}$ & 0.2842$^{***}$ \\
$\gamma=0.6$ & $(0.2,\,0.8)$ & 0.0057$^{***}$ & 0.0254$^{***}$ & 0.0301$^{***}$ & 0.0305$^{***}$ & 0.0305$^{***}$ \\
$\gamma=0.6$ & $(0.4,\,0.6)$ & -0.0000 & 0.0001 & 0.0065$^{***}$ & 0.0084$^{***}$ & 0.0097$^{***}$ \\
$\gamma=0.5$ (null) & $(0.2,\,0.8)$ & -0.0016$^{*}$ & -0.0023$^{***}$ & -0.0037$^{***}$ & -0.0034$^{***}$ & -0.0035$^{***}$ \\
$\gamma=0.5$ (null) & $(0.4,\,0.6)$ & 0.0006 & 0.0003 & -0.0054$^{***}$ & -0.0089$^{***}$ & -0.0102$^{***}$ \\
\bottomrule
\end{tabular}
\caption*{\small Sequential partition, $T=100$ draws, $I=100$ agents. $N$: observations per agent per draw. Stars: 10\% ($^*$), 5\% ($^{**}$), 1\% ($^{***}$).}
\end{table}


\clearpage

\begin{table}[ht]
\caption{Conditional Relevance of Demographic Characteristics}
\centering
\small
\begin{tabular}{lccccc}
\toprule
 & $N=5$ & $e$-GARP ($N=5$) & $N=1$ & $N=3$ & $e$-GARP ($N=1$) \\
\cmidrule(lr){2-2}\cmidrule(lr){3-3}\cmidrule(lr){4-4}\cmidrule(lr){5-5}\cmidrule(lr){6-6}
Variable & $D_G$ & $D_G$ & $D_G$ & $D_G$ & $D_G$ \\
\midrule
\emph{Family Size} \\
Low & -0.0059*** & +0.0010** & +0.0033*** & -0.0036*** & +0.0089*** \\
Mid & -0.0060*** & -0.0078*** & -0.0056*** & -0.0069*** & -0.0065*** \\
Large & -0.0384*** & -0.0432*** & -0.0320*** & -0.0378*** & -0.0321*** \\
\addlinespace
\emph{Income} \\
Low & -0.0224*** & -0.0199*** & -0.0124*** & -0.0199*** & -0.0043*** \\
Mid & -0.0016*** & -0.0020*** & -0.0025*** & -0.0019*** & -0.0021*** \\
High & -0.0226*** & -0.0265*** & -0.0190*** & -0.0222*** & -0.0213*** \\
\addlinespace
\emph{Age} \\
Young & -0.0245*** & -0.0246*** & -0.0232*** & -0.0244*** & -0.0144*** \\
Mid & -0.0043*** & -0.0050*** & -0.0039*** & -0.0041*** & -0.0044*** \\
Old & -0.0198*** & -0.0154*** & -0.0117*** & -0.0188*** & -0.0040*** \\
\addlinespace
\emph{Education} \\
Primary & -0.0047*** & -0.0038*** & -0.0034 & -0.0059*** & -0.0067** \\
High School & -0.0029*** & -0.0017*** & -0.0008 & -0.0032*** & +0.0016** \\
College & -0.0033*** & -0.0019*** & -0.0003 & -0.0038*** & +0.0026*** \\
\addlinespace
\bottomrule
\end{tabular}
\caption*{\small Conditional mean-difference statistic. Sequential partition, 480 households. $e$-GARP uses $e=0.95$ (5\% budget slack). $N$: observations per agent per draw. Stars: 10\% ($^*$), 5\% ($^{**}$), 1\% ($^{***}$).}
\label{tab:cond_demo}
\end{table}

\begin{table}[ht]
\caption{Conditional Permutation Test (Within-Strata Label Shuffling)}
\label{tab:cond_permutation}
\centering
\small
\begin{tabular}{lccc}
\toprule
Variable & $D_G$ (actual) & Perm.\ SD & $p$-value \\
\midrule
\emph{Family Size} \\
Low & +0.0033 & 0.0039 & 0.753 \\
Mid & -0.0056 & 0.0030 & 0.364 \\
Large & -0.0320 & 0.0076 & 0.114 \\
\emph{Income} \\
Low & -0.0124 & 0.0064 & 0.740 \\
Mid & -0.0025 & 0.0018 & 0.230 \\
High & -0.0190 & 0.0049 & 0.028$^{**}$ \\
\emph{Age} \\
Young & -0.0232 & 0.0079 & 0.336 \\
Mid & -0.0039 & 0.0032 & 0.341 \\
Old & -0.0117 & 0.0056 & 0.515 \\
\emph{Education} \\
Primary & -0.0034 & 0.0145 & 0.958 \\
High School & -0.0008 & 0.0026 & 0.798 \\
College & -0.0003 & 0.0031 & 0.924 \\
\bottomrule
\end{tabular}
\caption*{\small $1{,}000$ permutations of each demographic label \emph{within strata} defined by the other three demographic variables ($3^3 = 27$ background cells). Tests whether each variable contributes beyond what the remaining demographics predict. The within-stratum permutation distributions are discrete in finite samples, so the relationship between the actual $D_G$, the permutation SD, and the resulting $p$-value can differ from the Gaussian benchmark. After Benjamini--Hochberg FDR adjustment across the twelve dummies, no row survives at $q < 0.10$ (see Section~\ref{sec:application}). Stars: 10\% ($^*$), 5\% ($^{**}$), 1\% ($^{***}$).}
\end{table}

\begin{table}[ht]
\centering
\caption{Lottery application: partition kernel at $N\in\{1,5,10\}$, $T=500$, $I=686$ subjects.}
\label{tab:cpc18_summary}
\small
\begin{tabular}{lccc}
\toprule
& $N=1$ & $N=5$ & $N=10$ \\
\midrule
Mean partition kernel $\bar\pi^\Phi$ & $0.582$ & $0.092$ & $0.024$ \\
\quad sd                              & $0.071$ & $0.057$ & $0.062$ \\
\quad median                          & $0.584$ & $0.084$ & $0.016$ \\
\quad fraction $> 0.5$                & $0.876$ & $0.003$ & $0.004$ \\
\quad fraction $= 0$                  & $0.000$ & $0.001$ & $0.176$ \\
Joint-rationality gap $\bar\Delta_N = \bar\rho - \bar\pi^\Phi$ & $0.377$ & $0.867$ & $0.935$ \\
Centered $\lambda_1/\lambda_2$        & $4.14$  & $2.12$  & $1.18$ \\
\bottomrule
\end{tabular}
\caption*{\small Pairwise FSD-rationalisability $\bar\rho = 0.959$ (sd $0.058$, median $1$, $5$th percentile $0.840$). For comparison, scanner data at $N=1$: $\bar\pi^\Phi = 0.373$, $\bar\Delta = 0.618$, centered $\lambda_1/\lambda_2 = 3.2$. As $N$ grows, the partition imposes more cross-pair rationality constraints; cross-type pairs accumulate violations geometrically, so $\bar\pi^\Phi$ collapses by a factor of $24$ between $N=1$ and $N=10$. The drop in $\lambda_1/\lambda_2$ from $4.1$ to $1.2$ is the spectral signature of this sharpening.}
\end{table}


\clearpage
\section*{Figures}

\IfFileExists{cpc18_transitivity.pdf}{%
\begin{figure}[ht]
\centering
\includegraphics[width=0.85\textwidth]{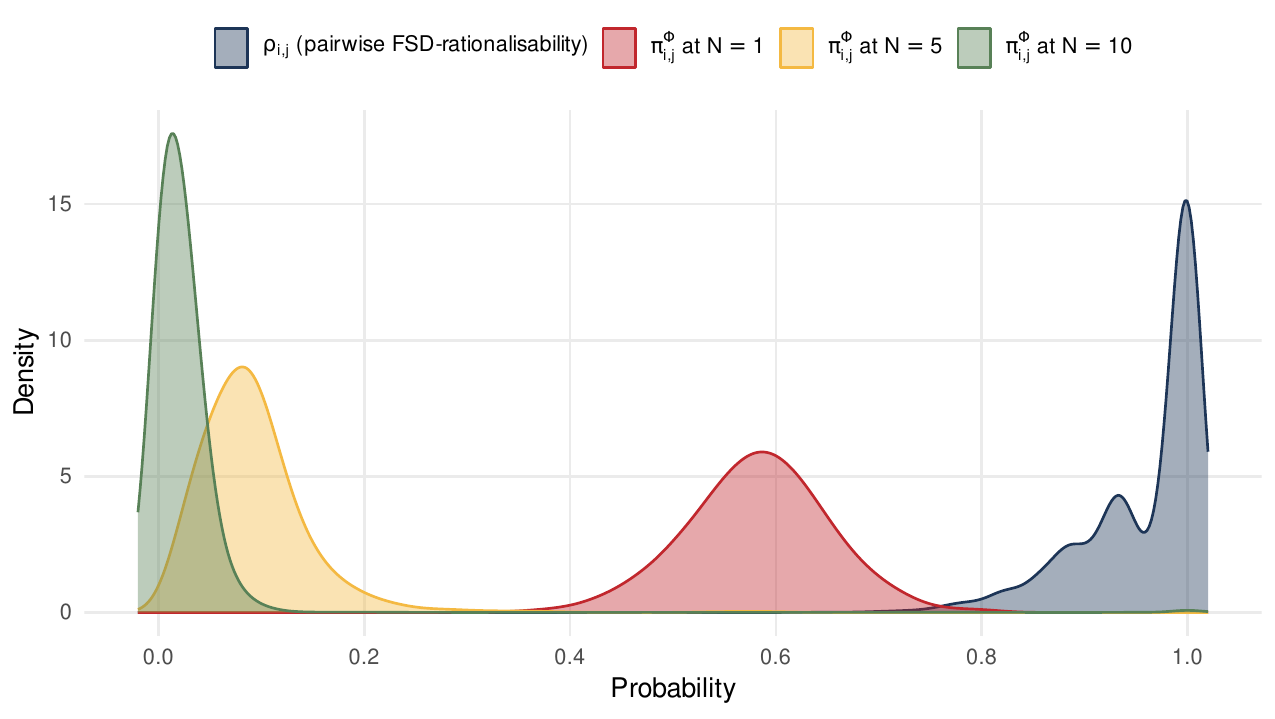}
\caption{Lottery application (CPC18 data). Densities across all $\binom{686}{2}$ subject pairs of the pairwise FSD-rationalisability $\rho_{i,j}$ (probability that one random binary choice from each subject is jointly rationalisable by some monotone utility) and the partition kernel entry $\pi_{i,j}^\Phi$ at $N\in\{1,5,10\}$. The pooled-rationality criterion is acyclicity of the joint directed preference graph augmented with FSD-implied edges. Sequential partition, $T=500$ Monte Carlo draws.}
\label{fig:cpc18_transitivity}
\end{figure}
}{}

\clearpage
\setcounter{page}{1}
\begin{center}
{\LARGE\bfseries Online Appendix}
\end{center}
\bigskip

\appendix
\renewcommand{\thetable}{\thesection.\arabic{table}}
\renewcommand{\thefigure}{\thesection.\arabic{figure}}
\renewcommand*{\theequation}{\thesection.\arabic{equation}}
\numberwithin{table}{section}
\numberwithin{figure}{section}
\numberwithin{equation}{section}
\numberwithin{theorem}{section}
\numberwithin{corollary}{section}
\numberwithin{lemma}{section}
\numberwithin{proposition}{section}
\numberwithin{definition}{section}
\numberwithin{assumption}{section}

\section{Cobb--Douglas illustration}\label{app:cobb_douglas}

This appendix illustrates the structural properties of the kernel using the two-good Cobb--Douglas specification employed in our simulation. Each agent maximises $u_i(q_1, q_2) = q_1^{\alpha_i} q_2^{1-\alpha_i}$ with $\alpha_i \in (0, 1)$, facing budget $p_{i,1} q_1 + p_{i,2} q_2 \leq 1$. The optimal bundle is $q_i^* = (\alpha_i / p_{i,1},\, (1-\alpha_i) / p_{i,2})$.

\paragraph{Pairwise GARP conditions.} For agents $i$ and $j$ with parameters $\alpha_i, \alpha_j$ and independently drawn prices from $[p_{\min}, p_{\max}]^2$, define price ratios $r_k = p_{i,k} / p_{j,k}$. The expenditure of agent $i$ on bundle $q_j$ is
\[
p_{i,1}\, q_{j,1} + p_{i,2}\, q_{j,2}
\;=\; \alpha_j\, \frac{p_{i,1}}{p_{j,1}} + (1 - \alpha_j)\, \frac{p_{i,2}}{p_{j,2}}
\;=\; \alpha_j\, r_1 + (1 - \alpha_j)\, r_2,
\]
and Cobb--Douglas exhausts the budget ($p_i \cdot q_i = 1$), so
\[
q_i\; R^0\; q_j \quad\Longleftrightarrow\quad \alpha_j\, r_1 + (1 - \alpha_j)\, r_2 \leq 1.
\]
By symmetry, $q_j\; P^0\; q_i \iff \alpha_i / r_1 + (1 - \alpha_i) / r_2 < 1$. A pairwise GARP violation requires both conditions simultaneously, yielding
\begin{align}
\text{Cycle 1:}\quad & \alpha_j\, r_1 + (1-\alpha_j)\, r_2 \leq 1
   \quad\text{and}\quad
   \tfrac{\alpha_i}{r_1} + \tfrac{1-\alpha_i}{r_2} < 1,
   \label{eq:cycle1}\\[4pt]
\text{Cycle 2:}\quad & \tfrac{\alpha_i}{r_1} + \tfrac{1-\alpha_i}{r_2} \leq 1
   \quad\text{and}\quad
   \alpha_j\, r_1 + (1-\alpha_j)\, r_2 < 1.
   \label{eq:cycle2}
\end{align}
Cycle~2 is obtained from Cycle~1 by exchanging $i$ and $j$.

\paragraph{No false detection.}

\begin{proposition}[No false detection]\label{prop:CD_no_false}
If $\alpha_i = \alpha_j$, neither cycle can hold for any $(r_1, r_2) \in \mathbb R_{++}^2$.
\end{proposition}

\begin{proof}
Set $\alpha = \alpha_i = \alpha_j$. By the Cauchy--Schwarz inequality,
\[
\bigl(\alpha\, r_1 + (1-\alpha)\, r_2\bigr)\!
\Bigl(\frac{\alpha}{r_1} + \frac{1-\alpha}{r_2}\Bigr)
\;\geq\;
\bigl(\alpha + (1-\alpha)\bigr)^2 = 1.
\]
Hence $\alpha\, r_1 + (1-\alpha)\, r_2 \leq 1$ implies $\alpha/r_1 + (1-\alpha)/r_2 \geq 1$, ruling out Cycle~1. Cycle~2 is symmetric.
\end{proof}

\paragraph{Distinct preferences.} When $\alpha_i \neq \alpha_j$, the Cauchy--Schwarz bound no longer binds because the two factors use different weights. The GARP violation probability $1 - \rho_{i,j}$ becomes strictly positive for price distributions with sufficient support and increases with $|\alpha_i - \alpha_j|$: when $\alpha_i \gg \alpha_j$, agent $i$ spends heavily on good~1 while agent $j$ focuses on good~2, and ``crossed'' price ratios ($r_1 > 1 > r_2$, i.e., agent~$i$ faces a higher price on her preferred good) bring both bundles within each other's budget sets, creating the overlap required for a GARP violation. Section~\ref{subsec:sim_calibration} computes $\rho_{i,j}$ numerically as a function of $|\alpha_i - \alpha_j|$ and confirms that the partition kernel $G$ reflects the latent type structure with same-type pairs exhibiting systematically higher similarity than different-type pairs.

\section{Approximate-optimisation extension}\label{app:eGARP}

This appendix extends the finite-type recovery theorem of the main text (Theorem~\ref{thm:finite_type_witness}) to approximate optimisation in the sense of \citet{halevy2018}, where each agent's individual dataset is allowed to satisfy $e$-GARP rather than exact GARP for some tolerance $e \in (0, 1]$.

\paragraph{Halevy et al.'s framework.}
A utility $u$ \emph{$e$-rationalises} a dataset $D$ if, for every observation $(p, q) \in D$ and every bundle $q'$ satisfying $e\, p \cdot q \geq p \cdot q'$, $u(q) \geq u(q')$. \citet[Theorem~1]{halevy2018} show that $D$ is $e$-rationalisable (by some continuous, monotone, concave utility) iff $D$ satisfies $e$-GARP, the cyclical consistency condition obtained from GARP by allowing budget slack at rate $1-e$. Setting $e = 1$ recovers Afriat's theorem. The sequential $e$-GARP-consistent partition rule is identical to the exact-GARP rule of Section~\ref{subsec:partition_rules} except that block compatibility is checked via $e$-GARP. Write $\pi_N^{\Phi, e}(\theta, \theta')$ for the population-averaged $e$-GARP partition kernel.

\paragraph{Finite-type recovery under $e$-GARP.}
The contrast-rank condition~\eqref{eq:CRC} of the main text has a direct $e$-GARP analogue: the type-level $e$-GARP kernel matrix $S_N^e := (\pi_N^{\Phi,e}(\theta_k, \theta_\ell))_{k,\ell=1}^K$ should be positive definite on the contrast subspace.

\begin{theorem}[Finite-type recovery under $e$-GARP]\label{thm:eGARP_witness}
Fix $e \in (0, 1]$. Suppose the population types are drawn from a $K$-point distribution $\mu = \sum_{k=1}^K w_k \delta_{\theta_k}$ and the type-level $e$-GARP kernel matrix $S_N^e$ satisfies
\[
z^\top S_N^e z > 0 \qquad \text{for every nonzero } z \in \mathbb R^K \text{ with } \mathbf 1_K^\top z = 0.
\]
Define the oracle $e$-GARP population kernel $\Pi^{\star,e} := E S_N^e E^\top$ via the type-membership matrix $E$. Then $H\Pi^{\star,e}H$ has rank exactly $K-1$, and its positive-eigenvalue eigenvectors are constant within each true type. Empirical recovery from a realised $e$-GARP kernel follows by Davis--Kahan perturbation under the analogue of Assumption~\ref{ass:empirical_oracle_kernel}, exactly as in Theorem~\ref{thm:finite_type_witness}.
\end{theorem}

\begin{proof}
Identical to that of Theorem~\ref{thm:finite_type_witness}, with $\Pi^{\star,e}$ replacing $\Pi_n^\star$ and $S_N^e$ replacing $S_N^\Phi$.
\end{proof}

\paragraph{Robustness check at $e = 0.95$.}
In the scanner data at $N=1$, the $e$-GARP partition kernel rises to mean $0.63$ from the exact-GARP $0.37$ as the $5\%$ slack unclogs block-level transitivity chains and produces larger blocks. The pairwise benchmark $\rho_e$ barely moves from $\rho$ (already $0.991$ at $e = 1$). The joint-rationality gap $\Delta^e = \rho_e - \pi^e$ shrinks with $e$ at finite $N$, but the demographic mean-difference structure --- which depends on the kernel's local geometry, not on the absolute size of the gap --- remains visible in the data.

\paragraph{The non-transitivity of $\sim_e$.}
A subtlety of approximate optimisation worth noting: the relation $i \sim_e j$ defined as ``$D_i \cup D_j$ satisfies $e$-GARP'' is in general not transitive at $e < 1$. Three agents may be pairwise $e$-compatible without being jointly $e$-rationalisable. This non-transitivity matters for asymptotic clustering claims (it can prevent same-class probability from reaching one as $N \to \infty$), but it does not affect the finite-type recovery result above: Theorem~\ref{thm:eGARP_witness} proceeds through the spectral structure of the $e$-GARP partition kernel and does not require a clustering limit.

\section{Proofs}\label{app:proofs}

\subsection*{Proofs of the permutation-test propositions}\label{app:perm_proofs}

\begin{proof}[Proof of Proposition~\ref{prop:perm_exact}]
Set $\sigma_0 := \mathrm{id}$ so that $D_G(\sigma_0 X) = D_G(X)$, and write $T_b := |D_G(\sigma_b X)|$ for $b = 0, 1, \ldots, B$. We show that $(T_0, T_1, \ldots, T_B)$ is exchangeable conditional on $G$, from which the bound on $\widehat p$ follows by a standard rank argument.

\emph{Exchangeability.} Fix any permutation $\tau$ of $\{0, 1, \ldots, B\}$. Under $H_0^{\mathrm{perm}}$, $X \mid G \stackrel{d}{=} \sigma X \mid G$ for any deterministic $\sigma \in S_{|\mathcal I|}$, hence for any random $\sigma$ independent of $X \mid G$. In particular, taking $\sigma = \sigma_{\tau(0)}$ (with $\sigma_{1:B}$ i.i.d.\ uniform and $\sigma_0 = \mathrm{id}$, both independent of $X \mid G$), set $X' := \sigma_{\tau(0)} X$, so $X' \mid G \stackrel{d}{=} X \mid G$. For $b \geq 1$, $\sigma_{\tau(b)} X = (\sigma_{\tau(b)} \sigma_{\tau(0)}^{-1}) X'$. Hence, conditional on $G$,
\[
\bigl(|D_G(\sigma_{\tau(0)} X)|,\, |D_G(\sigma_{\tau(1)} X)|,\, \ldots,\, |D_G(\sigma_{\tau(B)} X)|\bigr) \;\stackrel{d}{=}\; \bigl(|D_G(X')|,\, |D_G(\sigma_{\tau(1)} \sigma_{\tau(0)}^{-1} X')|,\, \ldots,\, |D_G(\sigma_{\tau(B)} \sigma_{\tau(0)}^{-1} X')|\bigr).
\]
Since $\{\sigma_b\}_{b=1}^B$ is an i.i.d.\ uniform sample on $S_{|\mathcal I|}$, the joint law of $\{\sigma_{\tau(b)} \sigma_{\tau(0)}^{-1}\}_{b=1}^B$ equals the joint law of $\{\sigma_b\}_{b=1}^B$ by right-invariance of the uniform distribution. Marginalising over $\sigma_{1:B}$ therefore gives
\[
(T_{\tau(0)}, \ldots, T_{\tau(B)}) \mid G \;\stackrel{d}{=}\; (T_0, T_1, \ldots, T_B) \mid G.
\]
This holds for every $\tau$, establishing conditional exchangeability of $(T_0, \ldots, T_B)$ given $G$.

\emph{Rank distribution.} Let $R := \#\{b \in \{0, 1, \ldots, B\} : T_b \geq T_0\}$ denote the rank of $T_0$ from the top among the $B+1$ statistics, with weak ties counted toward the larger rank. Exchangeability of $(T_b)_{b=0}^B$ implies that $R$ is uniformly distributed on $\{1, 2, \ldots, B+1\}$ in the absence of ties; with ties, $\Pr(R \leq r) \leq r/(B+1)$ for every $r \in \{1, \ldots, B+1\}$.

\emph{$p$-value bound.} The permutation $p$-value is $\widehat p = R/(B+1)$, so
\[
\Pr(\widehat p \leq \alpha) = \Pr\!\bigl(R \leq \alpha (B+1)\bigr) = \Pr\!\bigl(R \leq \lfloor \alpha (B+1) \rfloor\bigr) \leq \frac{\lfloor \alpha (B+1) \rfloor}{B+1} \leq \alpha. \qedhere
\]
\end{proof}

\begin{lemma}[Permutation variance bound]\label{lem:perm_var}
Let $G \in [0,1]^{n \times n}$ be a fixed symmetric array with zero diagonal, and let $X \in \{0,1\}^n$ satisfy $n_1/n \in [\delta, 1-\delta]$ for some $\delta \in (0, 1/2]$, where $n_x := \#\{i: X_i = x\}$. Let $\sigma$ be uniform on the symmetric group $S_n$ and define
\[
D_G(\sigma X) \;:=\; \overline G_{\mathrm w}(\sigma X) \;-\; \overline G_{\mathrm a}(\sigma X),
\]
where $\overline G_{\mathrm w}$ averages $G_{ij}$ over ordered pairs $(i,j)$, $i \neq j$, with $\sigma X_i = \sigma X_j$, and $\overline G_{\mathrm a}$ averages over pairs with $\sigma X_i \neq \sigma X_j$. Then
\[
\E_\sigma\!\bigl[D_G(\sigma X) \mid G, X\bigr] \;=\; 0, \qquad
\mathrm{Var}_\sigma\!\bigl[D_G(\sigma X) \mid G, X\bigr] \;\leq\; C(\delta)\,n^{-1},
\]
where $C(\delta)$ depends only on $\delta$.
\end{lemma}

\begin{proof}
Under uniform $\sigma$, $\Pr(\sigma X_i = \sigma X_j) = [n_0(n_0-1) + n_1(n_1-1)]/[n(n-1)]$ is the same for every ordered pair $i \neq j$, so both $\E_\sigma[\overline G_{\mathrm w}(\sigma X) \mid G, X]$ and $\E_\sigma[\overline G_{\mathrm a}(\sigma X) \mid G, X]$ equal the overall pair-mean of $G$; subtracting gives the mean claim.

For the variance, write
\[
D_G(\sigma X) \;=\; \sum_{i \neq j} G_{ij}\, c_{ij}(\sigma),
\qquad
c_{ij}(\sigma) \;:=\; \frac{I_{ij}(\sigma)}{|W|} \;-\; \frac{1 - I_{ij}(\sigma)}{|A|},
\]
with $I_{ij}(\sigma) := \mathbf 1\{\sigma X_i = \sigma X_j\}$, $|W| := n_1(n_1-1) + n_0(n_0-1)$, and $|A| := 2 n_0 n_1$. Under $n_1/n \in [\delta, 1-\delta]$ both $|W|$ and $|A|$ are $\Theta(n^2)$, so $|c_{ij}(\sigma)| \leq K(\delta) n^{-2}$ deterministically. Expanding,
\[
\mathrm{Var}_\sigma\!\bigl(D_G(\sigma X) \mid G, X\bigr) \;=\; \sum_{(i,j)}\sum_{(k,l)} G_{ij} G_{kl}\, \mathrm{Cov}_\sigma\!\bigl(c_{ij}(\sigma), c_{kl}(\sigma)\bigr),
\]
the sum running over ordered pairs $i \neq j$ and $k \neq l$. Partition by the overlap $|\{i,j\} \cap \{k,l\}| \in \{0,1,2\}$.

\emph{Two indices shared} ($\{i,j\} = \{k,l\}$). There are $O(n^2)$ such tuples and $|\mathrm{Cov}_\sigma(c_{ij}, c_{ij})| = \mathrm{Var}_\sigma(c_{ij}) \leq K(\delta)^2 n^{-4}$. Contribution: $O(n^{-2})$.

\emph{One index shared.} There are $O(n^3)$ such tuples. The covariance $\mathrm{Cov}_\sigma(I_{ij}, I_{il})$ for $j \neq l$ is a finite quantity bounded by a constant depending only on $\delta$ (its leading term is $p^3 + (1-p)^3 - (p^2 + (1-p)^2)^2$ with $p = n_1/n$, plus an $O(n^{-1})$ finite-population correction). Hence $|\mathrm{Cov}_\sigma(c_{ij}, c_{kl})| \leq K_1(\delta) n^{-4}$. Contribution: $O(n^{-1})$.

\emph{Disjoint pairs} ($\{i,j\} \cap \{k,l\} = \emptyset$). There are $O(n^4)$ such tuples. Under a uniform permutation, the values $\sigma X$ at four distinct positions form a uniform without-replacement sample from the multiset $\{X_1, \ldots, X_n\}$, so the joint law of indicators on disjoint pairs factorises up to a finite-population correction of order $n^{-1}$:
\[
\bigl|\mathrm{Cov}_\sigma(I_{ij}, I_{kl})\bigr| \;\leq\; K_2(\delta)\,n^{-1}.
\]
Therefore $|\mathrm{Cov}_\sigma(c_{ij}, c_{kl})| \leq K_2(\delta) n^{-5}$. Contribution: $O(n^{-1})$.

Summing the three regimes and using $|G_{ij}| \leq 1$ gives $\mathrm{Var}_\sigma(D_G(\sigma X) \mid G, X) \leq C(\delta) n^{-1}$.
\end{proof}

\begin{proof}[Proof of Proposition~\ref{prop:perm_consistent}]
Without loss of generality $\Delta^* > 0$. Write $n = |\mathcal I|$ and $h_{ij} = h_N^\Phi(\theta_i, \theta_j)$.

\emph{Oracle replacement.} By Assumption~\ref{ass:empirical_oracle_kernel}, the average pairwise error
\[
\bar e_n \;:=\; \binom{n}{2}^{-1} \sum_{i<j} |G_{ij} - h_{ij}| \;=\; o_p(1).
\]
Let $m_w = |\mathcal P_w|$ and $m_a = |\mathcal P_a|$. Since $X_i$ are i.i.d.\ $\mathrm{Bernoulli}(q)$ with $q \in (0,1)$, the weak law of large numbers gives $m_w/\binom{n}{2} \xrightarrow{p} q^2 + (1-q)^2 > 0$ and $m_a/\binom{n}{2} \xrightarrow{p} 2q(1-q) > 0$. Hence
\[
|D_G(X) - D^*(X)|
\;\leq\;
\frac{1}{m_w}\sum_{(i,j) \in \mathcal P_w} |G_{ij}-h_{ij}|
\,+\,
\frac{1}{m_a}\sum_{(i,j) \in \mathcal P_a} |G_{ij}-h_{ij}|
\;\leq\;
\left( \frac{\binom{n}{2}}{m_w} + \frac{\binom{n}{2}}{m_a} \right) \bar e_n
\;=\; O_p(1) \cdot o_p(1) \;=\; o_p(1),
\]
where the oracle statistic is
\[
D^*(X) \;:=\; \frac{1}{m_w}\sum_{(i,j) \in \mathcal P_w} h_{ij} \,-\, \frac{1}{m_a}\sum_{(i,j) \in \mathcal P_a} h_{ij}.
\]

\emph{U-statistic convergence of $D^*(X)$.} For any bounded symmetric kernel $f$ on $\Theta \times \{0,1\}$ define the order-two U-statistic
\[
U_n f \;:=\; \binom{n}{2}^{-1} \sum_{i<j} f\bigl((\theta_i,X_i),(\theta_j,X_j)\bigr).
\]
Then
\[
D^*(X) \;=\; \frac{U_n\{h_{12}\mathbf 1[X_1=X_2]\}}{U_n\{\mathbf 1[X_1=X_2]\}} \;-\; \frac{U_n\{h_{12}\mathbf 1[X_1\neq X_2]\}}{U_n\{\mathbf 1[X_1\neq X_2]\}}.
\]
By the law of large numbers for bounded U-statistics \citep{hoeffding1948u}, each numerator converges in probability to its expectation $\E[h_{12}\mathbf 1[X_1 = X_2]]$ (respectively $\E[h_{12}\mathbf 1[X_1 \neq X_2]]$), and the denominators converge to $q^2 + (1-q)^2 > 0$ and $2q(1-q) > 0$ respectively. By the continuous mapping theorem,
\[
D^*(X) \;\xrightarrow{p}\; \E[h_{12} \mid X_1 = X_2] - \E[h_{12} \mid X_1 \neq X_2] \;=\; \Delta^*.
\]
Combining with the oracle-replacement step gives $D_G(X) \xrightarrow{p} \Delta^* > 0$.

\emph{Permutation distribution.} Conditional on $(G, X)$ and on the event $\{n_0, n_1 \geq 1\}$, for any fixed pair $i \neq j$ the probability $\Pr_\sigma(\sigma X_i = \sigma X_j \mid X) = [n_0(n_0-1) + n_1(n_1-1)]/[n(n-1)]$ is the same across pairs, so $\E_\sigma[\overline G_{\mathrm w}(\sigma X) \mid G, X] = \E_\sigma[\overline G_{\mathrm a}(\sigma X) \mid G, X]$ equal the overall pair-mean of $G$; hence $\E_\sigma[D_G(\sigma X) \mid G, X] = 0$ exactly.

For the variance, choose any $\delta \in (0, \min\{q, 1-q\})$ and define the balance event
\[
\mathcal E_n \;:=\; \{ n_1/n \in [\delta, 1-\delta] \}.
\]
Since $n_1/n \xrightarrow{p} q$ with $q \in (\delta, 1-\delta)$, we have $\Pr(\mathcal E_n) \to 1$. On $\mathcal E_n$, Lemma~\ref{lem:perm_var} applies and yields
\[
\mathrm{Var}_\sigma\bigl(D_G(\sigma X) \mid G, X\bigr) \;\leq\; \frac{C(\delta)}{n}.
\]
By Chebyshev, on $\mathcal E_n$ and for every $\eta > 0$,
\[
\Pr_\sigma\!\bigl(|D_G(\sigma X)| \geq \eta \mid G, X\bigr) \;\leq\; \frac{C(\delta)}{n\,\eta^2} \;\xrightarrow[n \to \infty]{}\; 0.
\]

\emph{Combining.} Pick any $\alpha \in (0, 1)$ and set $\eta = \Delta^*/2$. With probability tending to one, both $\mathcal E_n$ and $\{|D_G(X)| \geq \eta\}$ hold. On their intersection,
\[
\frac{1}{B}\sum_{b=1}^{B} \mathbf 1\bigl\{|D_G(\sigma_b X)| \geq |D_G(X)|\bigr\}
\;\leq\;
\frac{1}{B}\sum_{b=1}^{B} \mathbf 1\bigl\{|D_G(\sigma_b X)| \geq \eta\bigr\}.
\]
The conditional expectation of the right-hand side given $(G, X)$ is bounded by $C(\delta)/(n\eta^2)$, so for $B = B(n, T) \to \infty$ the right-hand side is $o_p(1)$. The additive $1/(B+1)$ correction in the permutation $p$-value also vanishes. Therefore $\widehat p \xrightarrow{p} 0$ and $\Pr(\widehat p \leq \alpha) \to 1$. The case $\Delta^* < 0$ is identical after replacing $D_G$ by $-D_G$.
\end{proof}

\subsection*{Proof of the finite-type recovery theorem}\label{app:witness_proof}

\begin{proof}[Proof of Theorem~\ref{thm:finite_type_witness}]
Write $S = S_N^\Phi$. The proof has five steps.

\emph{Step 1: oracle positive eigenspace (item~(i) of the theorem).} Let $B_n := H_n E_n$ and $\mathcal C := \{z \in \mathbb R^K : \mathbf 1_K^\top z = 0\}$. Since all type shares are positive, $\mathrm{rank}(B_n) = K-1$, $B_n \mathbf 1_K = 0$, and $\mathrm{col}(B_n) = \mathcal V_n$. Moreover, $H_n \Pi_n^\star H_n = B_n S B_n^\top$. For any $v \in \mathcal V_n$, write $v = B_n c$ with $c \in \mathcal C$; this representation is without loss because adding a multiple of $\mathbf 1_K$ to $c$ does not change $B_n c$. Let $M_n := B_n^\top B_n = E_n^\top H_n E_n = \mathrm{diag}(n_k) - n^{-1}(n_k n_\ell)_{k,\ell}$ with $n_k = |I_k|$. A direct computation gives $c^\top M_n c = n \cdot \mathrm{Var}_p(c)$ where $p_k := n_k/n$, so $M_n \mathbf 1_K = 0$ and $\lambda_{\min}(M_n |_{\mathcal C}) \geq a n$ for some $a > 0$ and all $n$ large under the share lower bound. Hence $v \neq 0 \iff c \neq 0 \iff M_n c \neq 0$ in $\mathcal C$. Now
\[
\|v\|^2 \;=\; c^\top M_n c, \qquad v^\top H_n \Pi_n^\star H_n v \;=\; c^\top M_n S M_n c \;=\; (M_n c)^\top S (M_n c).
\]
Because $M_n c \in \mathcal C$, this quadratic form is strictly positive for every nonzero $v \in \mathcal V_n$ if and only if~\eqref{eq:CRC} holds. Hence the centred oracle kernel has exactly $K-1$ strictly positive eigenvalues iff~\eqref{eq:CRC}, with positive eigenspace $\mathcal V_n$; vectors in $\mathcal V_n$ are constant within each $I_k$ because $E_n$ is type-constant and centring subtracts the same scalar from all entries.

If~\eqref{eq:CRC} fails, there exists a nonzero $z \in \mathcal C$ with $z^\top S z \leq 0$. Since $S$ is positive semi-definite --- the type-level kernel is a Gram-style expectation by the same factorisation used in Theorem~\ref{thm:kernel}, applied to a population with one agent of each type --- the quadratic form's null space coincides with the operator null space, so $z^\top S z = 0$ implies $S z = 0$. Take any such $z^\star$, set $c^\star := M_n^{-1}|_{\mathcal C}\, z^\star \in \mathcal C$ (well-defined since $M_n$ is invertible on $\mathcal C$), and let $v^\star := B_n c^\star \in \mathcal V_n \setminus \{0\}$. Then $v^{\star\top} H_n \Pi_n^\star H_n v^\star = (M_n c^\star)^\top S (M_n c^\star) = z^{\star\top} S z^\star = 0$, so $v^\star$ is a nonzero vector in $\mathcal V_n$ lying in the kernel of $H_n \Pi_n^\star H_n$. In the simplest case $z^\star = e_k - e_\ell$, two distinct types collapse to the same point in the leading-eigenvector representation of the centred oracle kernel; for general null contrasts a weighted combination of types collapses, not necessarily a single pair.

\emph{Step 2: oracle eigengap scales linearly in $n$.} Using the representation $v = B_n c$ with $c \in \mathcal C$ from Step~1 and the identity $v^\top H_n \Pi_n^\star H_n v = (M_n c)^\top S (M_n c)$, we now bound the eigengap. The eigen-expansion of $c$ in the eigenbasis of $M_n |_{\mathcal C}$ gives
\[
\|M_n c\|^2 \;\geq\; \lambda_{\min}(M_n |_{\mathcal C}) \cdot c^\top M_n c \;\geq\; a n \cdot c^\top M_n c.
\]
Since $M_n c \in \mathcal C$, the contrast-rank condition~\eqref{eq:CRC} provides $b > 0$ with $z^\top S z \geq b \|z\|^2$ for all $z \in \mathcal C$, and therefore
\[
v^\top H_n \Pi_n^\star H_n v \;\geq\; b\, \|M_n c\|^2 \;\geq\; a b\, n\, c^\top M_n c \;=\; a b\, n\, \|v\|^2.
\]
The smallest positive eigenvalue $\lambda_n^\star := \lambda_{K-1}(H_n \Pi_n^\star H_n)$ therefore satisfies $\lambda_n^\star \geq a b\, n$ for $n$ large.

\emph{Step 3: operator-norm perturbation bound.} Let $R_{ij} := G_{ij} - S_{\tau(i),\tau(j)}$ for $i \neq j$ and $R_{ii} = 0$. Since $G_{ij}, S_{\tau(i),\tau(j)} \in [0,1]$, $|R_{ij}| \leq 1$ and hence $R_{ij}^2 \leq |R_{ij}|$. Specialising Assumption~\ref{ass:empirical_oracle_kernel} to $h_N^\Phi(\theta_k, \theta_\ell) = S_{k,\ell}$ gives $\sum_{i \neq j} |R_{ij}| = n(n-1) o_p(1)$, and therefore $\|R\|_F^2 \leq \sum_{i\neq j} |R_{ij}| = n(n-1) o_p(1)$, so $\|R\|_{\mathrm{op}} \leq \|R\|_F = o_p(n)$. The diagonal correction $D := \mathrm{diag}(1 - S_{\tau(i),\tau(i)})_i$ satisfies $\|D\|_{\mathrm{op}} \leq 1$. Hence $\|G - \Pi_n^\star\|_{\mathrm{op}} \leq \|R\|_{\mathrm{op}} + \|D\|_{\mathrm{op}} = o_p(n)$, and $\|H_n (G - \Pi_n^\star) H_n\|_{\mathrm{op}} = o_p(n)$ since $H_n$ does not increase operator norm.

\emph{Step 4: Davis--Kahan eigenspace consistency.} The oracle eigengap $\lambda_n^\star \geq ab\, n$ (Step~2) and the perturbation bound $o_p(n)$ (Step~3) feed into Davis--Kahan to give $\|\sin\Theta(\widehat{\mathcal V}_n, \mathcal V_n)\|_{\mathrm{op}} = o_p(1)$. Equivalently, there exists an orthogonal $R_n$ with $\|\widehat U_n - U_n^\star R_n\|_F = o_p(1)$, where $\widehat U_n$ is the leading $(K-1)$-dimensional eigenbasis of $H_n G H_n$.

\emph{Step 5: spectral clustering misclassification.} The rows of $U_n^\star$ take exactly $K$ distinct values, one per type, because $\mathcal V_n = \mathrm{col}(H_n E_n)$ consists of vectors constant within each true type (Step~1). Let $u_k \in \mathbb R^{K-1}$ denote the common row value for type $k$. Since $U_n^\star$ has $K-1$ orthonormal columns, $\sum_{k=1}^K n_k\|u_k\|^2 = \|U_n^\star\|_F^2 = K-1$; combined with $\liminf_n p_{k,n} > 0$ and the fact that the $K$ centres $\{u_k\}$ must span the $(K-1)$-dimensional contrast subspace, the minimum squared centroid separation $\min_{k \neq \ell}\|u_k - u_\ell\|^2$ is bounded below by $c(K)/n$ for some $c(K)>0$ depending only on $K$ and the type-share lower bound. Combining with the Frobenius eigenspace error from Step~4, the deterministic $k$-means misclassification bound \citep[Lemma~5.3]{LeiRinaldo2015} gives a misclassification rate $\leq C \|\widehat U_n - U_n^\star R_n\|_F^2 + o_p(1) = o_p(1)$.
\end{proof}

\begin{corollary}[Davis--Kahan rate]\label{cor:dk_rate}
Under the contrast-rank condition~\eqref{eq:CRC} and the type-share assumption $\liminf_n p_{k,n} > 0$, define the normalised oracle eigengap and the normalised perturbation
\[
\gamma_n \;:=\; n^{-1}\, \lambda_{K-1}\!\bigl(H_n \Pi_n^\star H_n\bigr),
\qquad
\eta_{nT} \;:=\; n^{-1}\,\bigl\| H_n(G - \Pi_n^\star) H_n \bigr\|_{\mathrm{op}}.
\]
Then $\gamma_n \geq ab$ for large $n$, where $a$ is the share lower bound from Step~2 of the proof and $b := \lambda_{\min}(S_N^\Phi |_{\mathcal C})$. Whenever $\eta_{nT} \leq \gamma_n/2$, there exists an orthogonal $R_n \in \mathcal O(K-1)$ such that
\[
\bigl\|\widehat U_n - U_n^\star R_n\bigr\|_F \;\leq\; C\sqrt{K-1}\,\frac{\eta_{nT}}{\gamma_n},
\]
for an absolute constant $C$ (from the Yu--Wang--Samworth form of Davis--Kahan). Under Assumption~\ref{ass:empirical_oracle_kernel}, $\eta_{nT} \leq \sqrt{a_{nT}} + n^{-1}$ where $a_{nT} := \frac{1}{n(n-1)}\sum_{i \neq j}\bigl|G_{ij} - \Pi^\star_{ij}\bigr|$ is the average off-diagonal oracle approximation error and the additive $n^{-1}$ term captures the bounded diagonal correction from Step~3 (since $G_{ii} = 1$ while $\Pi^\star_{ii} = S_{\tau(i),\tau(i)} \in [0,1]$). For the asymptotic statement, the off-diagonal term dominates and
\[
\bigl\|\widehat U_n - U_n^\star R_n\bigr\|_F \;=\; O_p\!\left(\frac{\sqrt{a_{nT}} + n^{-1}}{\gamma_n}\right).
\]
Combined with Lei--Rinaldo Lemma~5.3, the spectral-clustering misclassification rate is $O_p(a_{nT}/\gamma_n^2)$. The bounds follow directly from Steps~2--5 of the proof above: the eigengap lower bound $\lambda_n^\star \geq ab\,n$ from Step~2 gives $\gamma_n \geq ab$; the Frobenius--operator inequality $\|G - \Pi_n^\star\|_F^2 \leq n(n-1) a_{nT}$ from Step~3 gives $\eta_{nT} \leq \sqrt{a_{nT}}$ on $H_n$; Davis--Kahan in the Yu--Wang--Samworth form gives the eigenspace bound; and Lei--Rinaldo Lemma~5.3 gives the clustering bound.
\end{corollary}


\section{MILP Partition Procedure}\label{app:partition}

The MILP-based partition used in the robustness check (Table~\ref{tab:milp_comparison}) proceeds as follows. Fix a priority order $r:\mathcal I\to\{1,\dots,|\mathcal I|\}$ and set lexicographic weights $w_i=\varepsilon^{\,r(i)-1}$ with $\varepsilon\in(0,\tfrac12)$. For remainder set $\mathcal R\subseteq\mathcal I$, solve
\begin{align*}
\text{(OS)}\qquad
\max_{\{e_i\}_{i\in\mathcal R}} \;& C\sum_{i\in\mathcal R} e_i \;+\; \sum_{i\in\mathcal R} w_i\,e_i \\
\text{s.t. } \;& e_i\in\{0,1\}\ \forall i\in\mathcal R,\qquad
\bigl\{(p_i^{n_i^t},q_i^{n_i^t})\bigr\}_{i\in\mathcal R:\,e_i=1}\ \text{satisfies GARP,}
\end{align*}
where $C$ is large enough to prioritise cardinality over tie-breaking. The first term maximizes block size (the Houtman--Maks index; \citealp{houtman_maks}); the second resolves ties lexicographically. The GARP constraints are encoded via the MILP formulation of \citet{Demuynck2023} and \citet{HEUFER201587}. The partition is obtained by iterating: extract the solution block, remove those agents, and repeat on the remainder. Three tie-breaking variants (min, max, random priority) are used in the robustness check.

\section{Simulation and empirical-application details}\label{app:simulation_details}

This appendix collects the design, mechanism, and reading details of the simulation and scanner-application results that the body presents in summary form.

\paragraph{Permutation-test power calibration (Section~\ref{subsec:sim_diagnostics}).}
For each $\gamma \in \{0.5, 0.6, 0.8\}$, we run $R = 200$ outer replications. Each replication: (i)~draw $X_i \stackrel{\mathrm{iid}}{\sim} \mathrm{Bernoulli}(0.5)$; (ii)~draw types $\alpha_i \in \{0.2, 0.8\}$ symmetrically conditional on $X_i$: $\Pr(\alpha_i = 0.8 \mid X_i = 1) = \gamma$ and $\Pr(\alpha_i = 0.8 \mid X_i = 0) = 1-\gamma$ (so $\gamma=0.5$ is the symmetric null); (iii)~simulate $I = 100$ agents at $50$ price observations each on $[0.5, 5]^2$ and compute $G$ via the sequential partition at $N=1$, $T=100$; (iv)~run the two-sided permutation test with $B = 1{,}000$ shuffles. Table~\ref{tab:perm_power} reports the size and power. Under the null ($\gamma = 0.5$), the empirical rejection rate $0.055$ is well within the binomial sampling band around the nominal $0.05$, and the mean test statistic $\bar D_G = 0.0002$ is indistinguishable from zero. The two alternative rows trace the kernel's finite-sample power envelope: at $\gamma = 0.6$ the population-level mean-difference is $\bar D_G \approx 0.4$ percentage points and the test rejects with $35.5\%$ probability; at $\gamma = 0.8$, $\bar D_G \approx 4$ pp and the test rejects with probability one. 

\begin{table}[ht]
\centering
\caption{Finite-sample size and power of the permutation test.}
\label{tab:perm_power}
\small
\begin{tabular}{lcccc}
\toprule
$\gamma$ & $\bar D_G$ (sd) & rej.\ at $\alpha=0.05$ & rej.\ at $\alpha=0.10$ & interpretation \\
\midrule
$0.50$ & $\phantom{-}0.0002\ (0.0024)$ & $0.055$ & $0.105$ & null (size) \\
$0.60$ & $\phantom{-}0.0039\ (0.0050)$ & $0.355$ & $0.455$ & moderate alternative \\
$0.80$ & $\phantom{-}0.0405\ (0.0125)$ & $1.000$ & $1.000$ & strong alternative \\
\bottomrule
\end{tabular}
\caption*{\small Two-type Cobb--Douglas, $\alpha \in \{0.2, 0.8\}$. Each row averages $R = 200$ outer replications at $I = 100$, $T = 100$, $N = 1$, $B = 1{,}000$ permutations.}
\end{table}

\paragraph{$N$ spectrum mechanism (Section~\ref{subsec:sim_diagnostics}).}
Table~\ref{tab:N_spectrum} reports the discrimination ratio --- same-type to different-type mean similarity --- across $N \in \{1, 3, 5, 10, 25\}$. The ratio rises from $1.07$ at $N=1$ to over $500$ at $N=25$ because each additional within-agent observation imposes one more cross-pair rationality constraint when two agents are pooled, while same-type pooled data remains rationalisable by Afriat's theorem regardless of $N$. At $N=1$, per-pair discrimination is weak and the kernel functions as a population-level statistical object whose power comes from aggregation; at $N=25$, same-type agents are almost perfectly co-typed and the kernel approaches a block-diagonal structure aligned with latent types. Table~\ref{tab:small_sample} varies $(I, T)$ at fixed $N$ under $\gamma = 0.8$: with $I=25$ the method is fragile (small-population composition can reverse $D_G$); $I=50$ with $T \geq 100$ suffices for reliable detection of strong heterogeneity. The sequential rule with $I=500$ and $T=50$ runs in under two minutes, beyond MILP feasibility.

\begin{table}[ht]
\centering
\caption{Effect of $N$ on the kernel structure (sequential partition, 5 types, $I=100$, $T=100$)}
\label{tab:N_spectrum}
\small
\begin{tabular}{lccccc}
\toprule
$N$ & Mean $G_{i,j}$ & Mean block size & Singletons & Same-type $\bar G$ & Discrim.\ ratio \\
\midrule
1 & 0.523 & 25.3 & 11\% & 0.552 & 1.07 \\
3 & 0.218 & 15.0 & 7\% & 0.372 & 2.05 \\
5 & 0.158 & 11.8 & 7\% & 0.368 & 3.42 \\
10 & 0.119 & 9.5 & 7\% & 0.449 & 10.99 \\
25 & 0.173 & 16.1 & 0\% & 0.892 & 525.63 \\
\bottomrule
\end{tabular}
\caption*{\small ``Discrim.\ ratio'' is the ratio of same-type to different-type mean $G$; values above~$1$ indicate the kernel discriminates types. Larger $N$ provides sharper type discrimination.}
\end{table}

\begin{table}[ht]
\centering
\caption{Finite-sample properties: $D_G$ and $Z_G$ by population size ($I$) and number of draws ($T$)}
\label{tab:small_sample}
\small
\begin{tabular}{lcccccc}
\toprule
 & \multicolumn{2}{c}{$T=50$} & \multicolumn{2}{c}{$T=100$} & \multicolumn{2}{c}{$T=500$} \\
\cmidrule(lr){2-3}\cmidrule(lr){4-5}\cmidrule(lr){6-7}
 & $D_G$ & $Z_G$ & $D_G$ & $Z_G$ & $D_G$ & $Z_G$ \\
\midrule
$I=25$ & 0.019 & 1.63 & 0.078 & 7.02$^{***}$ & $-0.004$ & $-1.65$ \\
$I=50$ & 0.042 & 4.75$^{***}$ & 0.027 & 5.52$^{***}$ & 0.023 & 11.26$^{***}$ \\
$I=100$ & 0.028 & 5.44$^{***}$ & 0.035 & 7.45$^{***}$ & 0.022 & 15.15$^{***}$ \\
\bottomrule
\end{tabular}
\caption*{\small Two types $\alpha\in\{0.2,0.8\}$, $\gamma=0.8$, prices from $[0.5,5]^2$, $N=1$ observation per agent per draw, random tie-breaking. Stars: $^{***}$: $|Z|>2.576$.}
\end{table}

\begin{figure}[ht]
\centering
\includegraphics[width=0.78\textwidth]{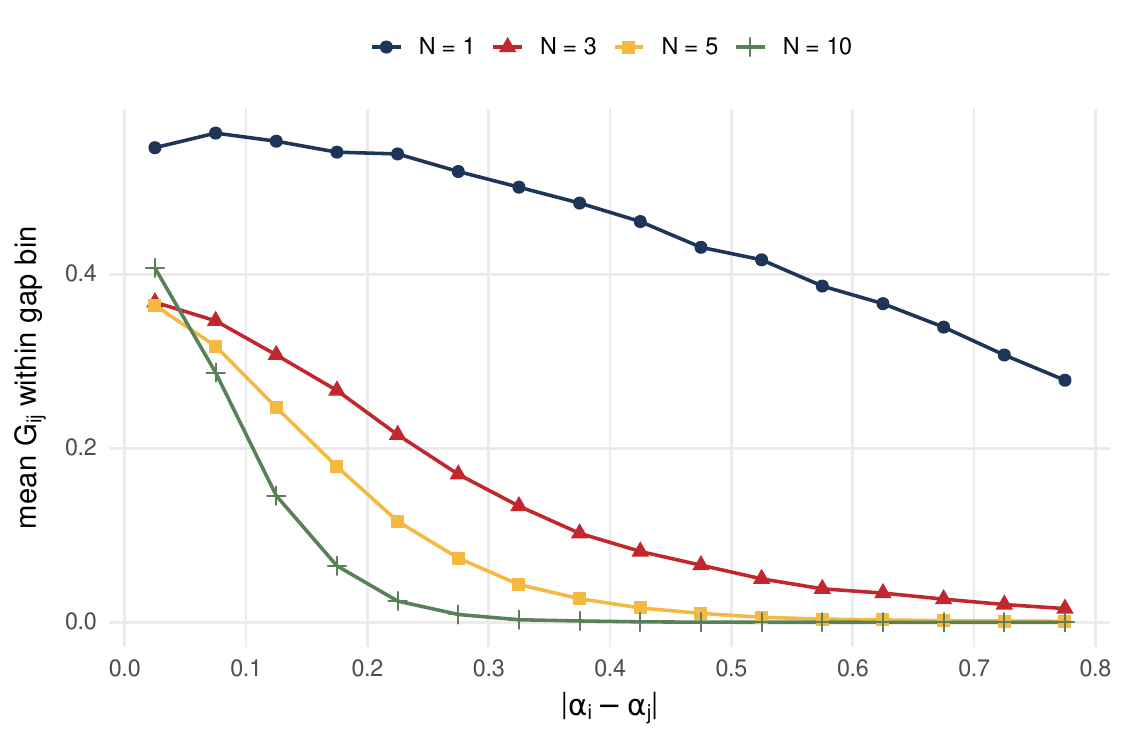}
\caption{Continuous-heterogeneity simulation. Mean kernel entry $G_{i,j}$, averaged within bins of the underlying parameter gap $|\alpha_i-\alpha_j|$, for $I=100$ agents with Cobb--Douglas parameters $\alpha_i \sim \mathrm{Uniform}(0.1, 0.9)$ and prices from $[0.5, 5]^2$. Sequential partition, $T=200$ draws. At small $N$ the kernel is a smooth decreasing function of parameter distance (bandwidth-like behaviour); at larger $N$ it collapses toward zero for most pairs (sharp type-indicator behaviour).}
\label{fig:continuous_heterogeneity}
\end{figure}

\paragraph{Continuous-heterogeneity numerics (Section~\ref{subsec:sim_diagnostics}).}
The simulation has $I=100$ agents with $\alpha_i \sim \mathrm{Uniform}(0.1, 0.9)$, prices on $[0.5, 5]^2$, at $N \in \{1, 3, 5, 10\}$ and $T = 200$. Figure~\ref{fig:continuous_heterogeneity} reports the mean kernel entry binned by the underlying parameter gap. At $N=1$ the mean kernel declines smoothly from $0.55$ at zero gap to $0.28$ at the largest gap. At $N=3$ the decay is sharper; at $N=10$, $67\%$ of pairs have $G < 0.1$. The kernel at small $N$ behaves like a smoothing bandwidth over the latent parameter space; at large $N$ it behaves like a sharp type-indicator, appropriate for finite-type structure but destructive of continuous similarity. 

\paragraph{Rule-dependence diagnostics (Section~\ref{subsec:sim_diagnostics}).}
Table~\ref{tab:milp_comparison} compares four partition rules --- the sequential rule and three MILP tie-breaks (min, max, random) --- on three metrics: Pearson correlation of off-diagonal kernel entries, Frobenius cosine of the full kernels, and absolute correlation of the leading eigenvectors. The four rules produce kernels that are moderately correlated at the pair level (Pearson $r = 0.73$ to $0.80$ across pairs of rules), very highly correlated globally (Frobenius cosine $\approx 0.99$), and well aligned on their leading eigenvectors ($|\mathrm{cor}(v_1)| = 0.82$ to $0.86$). Aggregate spectral and demographic objects are accordingly rule-stable. Specific contrast-level significances for smaller demographic effects can still shift across rules --- see Table~\ref{tab:robustness_demo} for the precise rule-sensitive cases in the scanner application --- but the kernels themselves are not pair-wise unrelated across rules.

\begin{table}[ht]
\centering
\caption{Robustness: Sequential vs.\ MILP partition (5 types, $I=100$, $T=100$)}
\label{tab:milp_comparison}
\small
\begin{tabular}{lccc}
\toprule
Comparison & Pearson $r$ & Frobenius cosine & $|\text{cor}(v_1)|$ \\
\midrule
Sequential vs MILP-min & 0.732 & 0.989 & 0.816 \\
Sequential vs MILP-max & 0.736 & 0.989 & 0.816 \\
Sequential vs MILP-random & 0.773 & 0.990 & 0.856 \\
MILP-min vs MILP-max & 0.733 & 0.988 & --- \\
MILP-min vs MILP-random & 0.798 & 0.990 & --- \\
MILP-max vs MILP-random & 0.763 & 0.989 & --- \\
\bottomrule
\end{tabular}
\caption*{\small Pearson $r$: correlation of off-diagonal $G_{i,j}$ entries. Frobenius cosine: global kernel similarity. $|\text{cor}(v_1)|$: absolute correlation of leading eigenvectors.}
\end{table}

\paragraph{Spectral decomposition of the scanner kernel (Section~\ref{sec:application}).}
The raw similarity matrix has unit diagonal and positive off-diagonal entries, so by Perron--Frobenius the leading raw eigenvalue is positive and corresponds to a positive eigenvector capturing the common-level component (rather than a heterogeneity dimension). The empirical ratio $\lambda_1/\lambda_2 = 181.8 / 3.2 \approx 56$ on the scanner kernel is a measured feature of this dataset; it reflects the strong level of co-typing across the population rather than a general consequence of positivity. The substantive point is that the raw spectrum is dominated by a level effect, not by dimensionality. The doubly centered kernel has eigenvalue ratio $\lambda_1/\lambda_2 = 3.2$. Applying the decomposition~\eqref{eq:DG_decomp} to the unconditional mean-difference statistic, $\lambda_1 D_1 / D_G = 103\%$ for large family size, $104\%$ for young age, $100\%$ for low income --- a direct consequence of the level dominance.

\section{Additional Figures and Tables}\label{app:additional_tables}

\begin{table}[ht]
\centering
\caption{Mean similarity $\bar G$ by type pair (sequential partition, 5 types)}
\label{tab:type_means}
\small
\begin{tabular}{lccccc}
\toprule
 & $\alpha=0.20$ & $\alpha=0.35$ & $\alpha=0.50$ & $\alpha=0.65$ & $\alpha=0.80$ \\
\midrule
$\alpha=0.20$ & \textbf{0.445} & 0.505 & 0.506 & 0.458 & 0.374 \\
$\alpha=0.35$ & 0.505 & \textbf{0.622} & 0.645 & 0.578 & 0.458 \\
$\alpha=0.50$ & 0.506 & 0.645 & \textbf{0.696} & 0.631 & 0.505 \\
$\alpha=0.65$ & 0.458 & 0.578 & 0.631 & \textbf{0.592} & 0.495 \\
$\alpha=0.80$ & 0.374 & 0.458 & 0.505 & 0.495 & \textbf{0.445} \\
\bottomrule
\end{tabular}
\caption*{\small Each entry averages $G_{i,j}$ over all pairs with the indicated type combination. Bold: same-type (diagonal). $T=100$, $I=100$ (20 per type), prices from $[0.5,5]^2$.}
\end{table}

\begin{table}[ht]
\centering
\begin{threeparttable}
\centering
\begin{tabular}{lc}
\toprule
Variable & Number of Households\\
\hline
\emph{Family Size} & \\
Low & 228\\
Mid & 187\\
Large & 65\\
\emph{Income} & \\
Low & 139\\
Mid & 200\\
High & 141\\
\emph{Age} & \\
Young & 122\\
Mid & 201\\
Old & 157\\
\emph{Education} & \\
Primary Education & 28\\
High School & 197\\
College & 255\\
\addlinespace
Observations & 480\\
\bottomrule
\end{tabular}
\begin{tablenotes}
\small
\item Middle-aged: average spouse age 30--65. Old: above 65. Mid-size: 3--4 members. Large: more than 4. Low income: below \$20{,}000. Mid: \$20{,}000--\$45{,}000. High: above \$45{,}000.
\end{tablenotes}
\end{threeparttable}
\caption{Sociodemographic Variables}
\label{tab:summary_stat_scanner_data}
\end{table}

\begin{table}[t]
\centering
\caption{Robustness across the Afriat efficiency path ($N=5$, $T=500$, scanner data, $480$ households)}
\label{tab:egarp_path}
\small
\begin{threeparttable}
\begin{tabular}{l*{5}{c}}
\toprule
& \multicolumn{5}{c}{Afriat efficiency tolerance $e$} \\
\cmidrule(lr){2-6}
& 1.00 & 0.99 & 0.97 & 0.95 & 0.90 \\
\midrule
Mean kernel $\bar\pi$            & 0.037 & 0.047 & 0.072 & 0.111 & 0.281 \\
Mean pairwise $\bar\rho$         & 0.691 & 0.752 & 0.837 & 0.898 & 0.965 \\
Transitivity gap $\bar\Delta$    & 0.654 & 0.705 & 0.765 & 0.787 & 0.685 \\
$\lambda_1/\lambda_2$ (centred)  & 2.61 & 3.07 & 4.35 & 5.83 & 9.90 \\
\addlinespace
\multicolumn{6}{l}{\emph{Conditional $D_G$ (largest four headline effects)}} \\
\quad Large family size       & $-$0.038$^{***}$ & $-$0.038$^{***}$ & $-$0.041$^{***}$ & $-$0.043$^{***}$ & $-$0.052$^{***}$ \\
\quad Old age                 & $-$0.020$^{***}$ & $-$0.019$^{***}$ & $-$0.017$^{***}$ & $-$0.015$^{***}$ & $-$0.006$^{***}$ \\
\quad High income             & $-$0.022$^{***}$ & $-$0.023$^{***}$ & $-$0.023$^{***}$ & $-$0.026$^{***}$ & $-$0.034$^{***}$ \\
\quad College                 & $-$0.004$^{***}$ & $-$0.003$^{***}$ & $-$0.003$^{***}$ & $-$0.002$^{***}$ & $+$0.001 \\
\bottomrule
\end{tabular}
\begin{tablenotes}\footnotesize
\item Notes. The Afriat efficiency tolerance $e$ relaxes exact GARP via \citet{halevy2018}: a dataset is $e$-rationalisable if it satisfies $e$-GARP, where the cyclical-consistency condition allows budget slack at rate $1-e$. The mean kernel rises with relaxation as more pools become $e$-GARP-consistent; the pairwise benchmark $\bar\rho$ also rises. The joint-rationality gap $\bar\Delta = \bar\rho - \bar\pi$ peaks around $e=0.95$ and contracts at $e=0.90$ as the pairwise benchmark approaches saturation. Of the four headline conditional mean-difference statistics, large family size, old age, and high income retain sign across the path (with large family size and high income strengthening, old age attenuating); College is small throughout and loses significance at $e=0.90$, where it crosses zero (from $-$0.004 to $+$0.001). The ratio $\lambda_1/\lambda_2$ of the doubly centred kernel rises monotonically with relaxation. Stars on the conditional $D_G$ entries: 10\% ($^*$), 5\% ($^{**}$), 1\% ($^{***}$).
\end{tablenotes}
\end{threeparttable}
\end{table}

\begin{table}[ht]
\caption{Permutation Validation of Demographic Associations}
\label{tab:permutation}
\centering
\small
\begin{tabular}{lccc}
\toprule
Variable & $D_G$ (actual) & Perm.\ SD & $p$-value \\
\midrule
\emph{Family Size} \\
Low & +0.0033 & 0.0004 & $<0.001$$^{***}$ \\
Mid & -0.0049 & 0.0017 & 0.006$^{***}$ \\
Large & -0.0199 & 0.0051 & $<0.001$$^{***}$ \\
\emph{Income} \\
Low & +0.0098 & 0.0029 & $<0.001$$^{***}$ \\
Mid & -0.0013 & 0.0012 & 0.281 \\
High & -0.0063 & 0.0028 & 0.029$^{**}$ \\
\emph{Age} \\
Young & -0.0174 & 0.0035 & $<0.001$$^{***}$ \\
Mid & +0.0012 & 0.0012 & 0.345 \\
Old & +0.0098 & 0.0025 & $<0.001$$^{***}$ \\
\emph{Education} \\
Primary & +0.0085 & 0.0077 & 0.271 \\
High School & -0.0001 & 0.0013 & 0.943 \\
College & +0.0001 & 0.0005 & 0.759 \\
\bottomrule
\end{tabular}
\caption*{\small $1{,}000$ random permutations of demographic labels on the same kernel $G$. Perm.\ SD: standard deviation of $D_G$ under the permutation null. Stars: 10\% ($^*$), 5\% ($^{**}$), 1\% ($^{***}$).}
\end{table}

\begin{table}[ht]
\caption{Partition-Rule Robustness: Conditional Tests on 300-Household Subsample}
\label{tab:robustness_demo}
\centering
\small
\begin{tabular}{llcccc}
\toprule
 & & \multicolumn{4}{c}{$D_G(k)$} \\
\cmidrule(lr){3-6}
 & Variable & Sequential & MILP-min & MILP-max & MILP-random \\
\midrule
\emph{Family Size} \\
& Low & +0.0025** & +0.0240*** & +0.0186*** & +0.0165*** \\
& Mid & -0.0044*** & -0.0023 & -0.0028 & -0.0028 \\
& Large & -0.0339*** & -0.0334*** & -0.0295*** & -0.0249*** \\
\emph{Income} \\
& Low & -0.0279*** & +0.0013 & -0.0049 & -0.0002 \\
& Mid & -0.0015** & -0.0007 & -0.0006 & -0.0006 \\
& High & -0.0149*** & -0.0088** & -0.0030 & -0.0052 \\
\emph{Age} \\
& Young & -0.0177*** & -0.0125*** & -0.0136*** & -0.0091*** \\
& Mid & -0.0043*** & -0.0081*** & -0.0040* & -0.0091*** \\
& Old & -0.0342*** & -0.0007 & -0.0025 & -0.0031 \\
\emph{Education} \\
& Primary & +0.0068** & +0.0172*** & -0.0027 & +0.0090 \\
& High School & -0.0066*** & -0.0021 & -0.0017 & -0.0060*** \\
& College & -0.0052*** & +0.0035 & +0.0084*** & +0.0057** \\
\bottomrule
\end{tabular}
\caption*{\small Conditional mean-difference statistic on the same 300-household subsample. Column 1: sequential partition ($T=500$). Columns 2--4: MILP partition with min, max, and random tie-breaking ($T=100$). Stars: 10\% ($^*$), 5\% ($^{**}$), 1\% ($^{***}$).}
\end{table}

\end{document}